\documentclass[a4paper,11pt]{article}
\usepackage{pos}

\renewcommand{\logo}{\relax}

\usepackage{subcaption}
\usepackage[T1]{fontenc}
\usepackage{float}
\usepackage{latexsym}
\usepackage{t1enc}
\usepackage{indentfirst}
\usepackage{pb-diagram}
\usepackage{graphicx}
\usepackage{enumerate}
\usepackage{color}
\usepackage[all]{xy}
\usepackage{hyperref}
\hypersetup{colorlinks=false,pdfborderstyle={/S/U/W 0}}
\usepackage{url}
\usepackage{upgreek}

\theoremstyle{plain}
\newtheorem{thm}{Theorem}[section]             

\newtheorem{prop}[thm]{Proposition}

\theoremstyle{definition}
\newtheorem{definition}[thm]{Definition}

\theoremstyle{remark}

\newcommand{\be}{\begin{equation*}}
\newcommand{\ee}{\end{equation*}}
\newcommand{\ben}{\begin{equation}}
\newcommand{\een}{\end{equation}}
\newcommand{\beqa}{\begin{eqnarray*}}
\newcommand{\eeqa}{\end{eqnarray*}}
\newcommand{\beqan}{\begin{eqnarray}}
\newcommand{\eeqan}{\end{eqnarray}}
\newcommand{\nn}{\nonumber}

\def\i{\mathbf{i}}
\def \const{\mathrm{const}}

\def\R{\mathbb{R}}

\def\Hess{\mathrm{Hess}}

\def\Crit{\mathrm{Crit}}

\def\hV{\widehat{V}}

\def\const{\mathrm{const}}

\newcommand{\pd}{\partial}
\def\dd{\mathrm{d}}

\newcommand{\str}{{\mathrm str}}

\newcommand{\sign}{\mathrm{sign}}

\def\cG{\mathcal{G}}
\def\cH{\mathcal{H}}

\def\cM{\mathcal{M}}

\def\cV{\mathcal{V}}

\def\IR{\mathrm{IR}}

\def\rmax{\mathrm{max}}

\def\rA{\mathrm{A}}

\def\rS{\mathrm{S}}

\def\e{{\bf e}}

\def\c{{\bf c}}

\newcommand{\eqdef}{\stackrel{{\rm def.}}{=}}

\def\grad{\mathrm{grad}}

\def\O{\mathrm{O}}

\def\hSigma{\widehat{\Sigma}}
\def\hV{\widehat{V}}

\def\hPhi{{\hat \Phi}}

\def\bvert{\big{\vert}}

\def\rM{\mathrm{M}}

\newcommand{\fourpartdef}[8]
{
	\left\{
	\begin{array}{ll}
		#1 & \mbox{if} #2 \\
		#3 & \mbox{if} #4 \\
		#5 & \mbox{if} #6\\
                #7 & \mbox{if} #8
	\end{array}
	\right.
}

\newcommand{\twopartdef}[4]
{
	\left\{
	\begin{array}{ll}
		#1 & \mbox{if } #2 \\
		#3 & \mbox{if } #4
	\end{array}
	\right.
}

\title{Infrared behavior in tame hyperbolizable two-field models}
\ShortTitle{IR behavior in tame hyperbolizable two-field models}

\author*{Elena Mirela Babalic}
\author{Calin Iuliu Lazaroiu}

\affiliation{Horia Hulubei National Institute of Physics and Nuclear Engineering, Department of Physics,\\
  Reactorului 30, Bucharest-Magurele, 077125, Romania}

\emailAdd{mbabalic@theory.nipne.ro}
\emailAdd{lcalin@theory.nipne.ro}

\abstract{We discuss the behavior of cosmological curves and their
  first order infrared approximants near critical ends of the scalar
  manifold $\Sigma$ and near interior critical points of the scalar
  potential for tame hyperbolizable two-field cosmological models by
  determining the universal forms of the asymptotic gradient flow of
  the classical effective potential with respect to the uniformizing
  metric near all these points and ends.  We compare the asymptotic
  behavior of gradient flow curves with numerical results for
  cosmological curves.}

\FullConference{%
  11th International Conference of the Balkan Physical Union (BPU11),\\
  28 August - 1 September 2022\\
  Belgrade, Serbia
}

\begin{document}
\maketitle

\section{Introduction}

Two-field cosmological models provide the simplest testing ground for
multifield cosmological dynamics. The latter are of great importance for
connecting cosmology with fundamental theories of gravity and matter,
since the effective description of generic string and M-theory
compactifications contains many moduli fields. In particular,
multifield models are crucial in cosmological applications of the
swampland program \cite{V, OV, BCV, BCMV}, as pointed out for example
in \cite{AP, OOSV,GK}, and may also allow for a unified description of
inflation, dark matter and dark energy \cite{AL}.

A two-field cosmological model is parameterized by a connected
borderless smooth surface $\Sigma$ (the target manifold of the scalar
fields) endowed with a Riemannian metric $\cG$ and a scalar potential
$\Phi$ which is a smooth function defined on $\Sigma$. We assume that
$\Phi$ is positive everywhere. In \cite{ren}, we used a dynamical RG 
(renormalization group) flow analysis and the uniformization theorem of Poincar\'e to show
that two-field models whose scalar field metric $\cG$ has constant
Gaussian curvature $K$ equal to $-1$, $0$ or $+1$ give distinguished
representatives for the IR (infrared) universality classes of all two-field
cosmological models. {\em Hyperbolizable two-field models} (which are
defined as those models for which $K=-1$) comprise all two-field
models whose target is of general type as well as those models whose
target is exceptional (i.e diffeomorphic with $\R^2$, the annulus
$\rA^2$ or the M\"obius strip $\rM^2$) and for which the metric belongs
to a hyperbolizable conformal class. The uniformized form of a
hyperbolizable model is a {\em two-field generalized
  $\alpha$-attractor model} in the sense of \cite{genalpha}. Some
aspects of such models were investigated previously in
\cite{elem,modular, Noether1, Noether2, Hesse, Lilia1,Lilia2} (see
\cite{unif,Nis,Tim19, LiliaRev} for brief reviews).

The infrared behavior of a tractable class of hyperbolizable two-field
models was studied in \cite{grad}, work which we summarize here
together with a brief announcement of further results. We will assume
that the target manifold $\Sigma$ is oriented and topologically finite
in the sense that it has finitely-generated fundamental group. When
$\Sigma$ is non-compact, this condition insures that it has a finite
number of Freudenthal ends \cite{Freudenthal1} and that its end
compactification $\hSigma$ is a smooth and oriented compact
surface. Thus $\Sigma$ is recovered from $\hSigma$ by removing a
finite number of points.  We also assume that the scalar potential
$\Phi$ admits a smooth extension $\hPhi$ to $\hSigma$ which is a
strictly-positive Morse function defined on $\hSigma$. A two-field
cosmological model is called {\em tame} when these conditions are
satisfied. Thus tame hyperbolizable two-field cosmological models are
those classical two-field models whose scalar manifold is a
connected, oriented and topologically finite hyperbolizable Riemann
surface $(\Sigma,\cG)$ and whose scalar potential $\Phi$ admits a positive
and Morse extension to the end compactification of $\Sigma$.

\paragraph{Notations and conventions.}

All surfaces $\Sigma$ considered here are connected,
smooth, Hausdorff and paracompact. If $V$ is a smooth real-valued
function defined on $\Sigma$, we denote by:
\be
\Crit V\eqdef \{c\in \Sigma| (\dd V)(c)=0\}
\ee
the set of its critical points. For any $c\in \Crit V$, we denote
by $\Hess(V)(c)$ the Hessian of $V$ at $c$,
which is a well-defined and coordinate independent symmetric bilinear
form on the tangent space $T_c\Sigma$. Given a metric $\cG$ on
$\Sigma$, we define the covariant Hessian tensor of $V$ relative to $\cG$ by:
\be
\Hess_\cG(V)\eqdef \nabla\dd V~~,
\ee
where
$\nabla$ is the Levi-Civita connection of $\cG$. This symmetric tensor
has the following local expression in coordinates
$(x^1,x^2)$ on $\Sigma$:
\be
\Hess_\cG(V)=(\pd_i\pd_j-\Gamma^k_{ij}(x)\pd_k)V \dd x^i\otimes \dd
x^j~~,
\ee
where $\Gamma^k_{ij}(x)$ are the Christoffel symbols of $\cG$.  Recall
that a critical point $c$ of $V$ is called {\em nondegenerate} if
$\Hess_\cG(V)(c)$ is a non-degenerate bilinear form. When $V$ is a
Morse function (i.e. has only non-degenerate critical points), the set
$\Crit V$ is discrete.  We denote by $\hSigma$ the end
compactification of $\Sigma$, which is a compact Hausdorff topological
space containing $\Sigma$. In the topologically finite case, the
surface $\Sigma$ has a finite number of Freudenthal ends and $\hSigma$
is a smooth compact surface. In this situation, we say that $V$ is
{\em globally well-behaved} on $\Sigma$ if it admits a smooth
extension $\hV$ to $\hSigma$.  A metric $\cG$ on $\Sigma$ is called
{\em hyperbolic} if it is complete and of constant Gaussian curvature
$K=-1$.

\section{Hyperbolizable two-field models}

Let us recall the global description of two-field cosmological
models through a second order geometric ODE and the first order
infrared approximation introduced in \cite{ren}. Such a model is
parameterized by the rescaled Planck mass $M_0\eqdef
M\sqrt{\frac{2}{3}}$ (where $M$ is the reduced Planck mass) and by its
{\em scalar triple} $(\Sigma,\cG,\Phi)$, where $\Sigma$ is the target
manifold for the scalar fields (a generally non-compact borderless
connected surface), $\cG$ is the scalar field metric and $\Phi$ is the
scalar potential. To ensure conservation of energy, one requires that
$\cG$ is complete. For simplicity, we also assume that $\Phi$ is strictly positive.

When neglecting fluctuations, the scalar field $\varphi:\R\rightarrow
\Sigma$ coupled to the metric of a
Friedmann-Lema\^itre-Robertson-Walker (FLRW) space satisfies the {\em cosmological
  equation} (see (1.4) in \cite{grad}):
\ben
\label{cosm}
\nabla_t \dot{\varphi}(t)+\frac{1}{M_0} \left[||\dot{\varphi}(t)||_\cG^2+2\Phi(\varphi(t))\right]^{1/2}\dot{\varphi}(t)+
(\grad_{\cG} \Phi)(\varphi(t))=0~~.
\een

\begin{prop}
The IR behavior (in the sense of \cite{ren}) of the cosmological flow
of a two-field model with scalar triple $(\Sigma,\cG,\Phi)$ and
rescaled Planck mass $M_0$ is described by the gradient flow of the
scalar triple $(\Sigma,G,V)$, where $G$ is the uniformizing metric of
$\cG$ and $V\eqdef M_0\sqrt{2\Phi}$ is the classical effective scalar
potential of the model.
\end{prop}

\noindent Hence the IR behavior is described by the gradient flow equation:
\ben
\label{gradV}
\dot\varphi_{\IR}(t) + (\grad_{G} {V})(\varphi_{\IR}(t))=0~.
\een
The first order IR approximants of cosmological orbits for the model
$(M_0,\Sigma,\cG,\Phi)$ coincide with those of the model
$(M_0,\Sigma,G,\Phi)$. Moreover, these approximants coincide with the
gradient flow orbits of $(\Sigma,G,V)$. In particular, the IR
universality classes defined in \cite{ren} depend only on the scalar
triple $(\Sigma,G,V)$. This allows for systematic studies of two-field
cosmological models belonging to a fixed IR universality class by
using the infrared expansion of cosmological curves, the first order
of which is given by the gradient flow of the classical effective
potential $V$ on the geometrically finite hyperbolic surface
$(\Sigma,G)$. Since the future limit points of cosmological curves and
of the gradient flow curves of $(\Sigma,G,V)$ are critical points of
$\Phi$ or Freudenthal ends of $\Sigma$, the asymptotic behavior of
such curves for late cosmological times is determined by the form of
$G$ and $V$ near such points.

\subsection{The hyperbolic metric $G$ in the vicinity of an end}

\noindent In this subsection, we recall the form of the hyperbolic metric $G$ in a canonical
vicinity of an end and extract its asymptotic behavior near each type
of end.

Any end $\e$ of $\Sigma$ admits an open neighborhood $U_\e\subset
\hSigma$ diffeomorphic with a disk such that there exist {\em
  semigeodesic polar coordinates} $(r,\theta)\in \R_{>0}\times \rS^1$
defined on $\dot{U}_\e\eqdef U_\e\setminus \{\e\}\subset \Sigma$ in
which the metric $G$ has the canonical form:
\be
\label{emetric}
\dd s_G^2|_{\dot{U}_\e}=\dd r^2+f_\e(r)\dd \theta^2~~,
\ee

 \be
\label{fe}
f_\e(r)= \fourpartdef{\sinh^2(r)}{~\e=\mathrm{plane~end}}
{\frac{1}{(2\pi)^2}e^{2r}}{~\e=\mathrm{horn~end}}
{\frac{\ell^2}{(2\pi)^2}\cosh^2(r)}{~\e=\mathrm{funnel~end~of~circumference}~\ell>0}{\frac{1}{(2\pi)^2}e^{-2r}}{~\e=\mathrm{cusp~end}}~~.
\ee
The end corresponds to $r\!\rightarrow \!\infty$. 
Setting $\omega\!\eqdef \!\frac{1}{r}$, the metric in {\em canonical polar coordinates} $(\omega,\theta)$ is:
 \be
\label{eomegametric}
\dd s_G^2|_{\dot{U}_\e}=\frac{\dd \omega^2}{\omega^4}+f_\e(1/\omega)\dd \theta^2~~,
\ee
where:
\ben
\label{feas}
f_\e(1/\omega)= {\tilde c}_\e e^{\frac{2\varepsilon_\e}{\omega}}\left[1+\O\left(e^{-\frac{2}{\omega}}\right)\right]~~\mathrm{for}~~\omega\to 0~~,
\een
with:
\beqa
&&{\tilde c}_\e=\fourpartdef{\frac{1}{4}}{~\e=\mathrm{plane~end}}
{\frac{1}{(2\pi)^2}}{~\e=\mathrm{horn~end}}
{\frac{\ell^2}{(4\pi)^2}}{~\e=\mathrm{funnel~end~of~circumference}~\ell>0}{\frac{1}{(2\pi)^2}}{~\e=\mathrm{cusp~end}}\\
&&\varepsilon_\e=\twopartdef{+1}{~\e=\mathrm{flaring~(i.e.~plane,~horn~or~funnel)~end}}{-1}{~\e=\mathrm{cusp~end}}
\eeqa
The term $\O\left(e^{-\frac{2}{\omega}}\right)$ in \eqref{feas}
vanishes identically when $\e$ is a cusp or horn end. In particular,
the constants ${\tilde c}_\e$ and $\epsilon_\e$ determine the leading
asymptotic behavior of the hyperbolic metric $G$ near $\e$.

The gradient flow equations of $(\dot{U}_\e,G|_{\dot{U}_\e},V|_{\dot{U}_\e})$ read:
\beqan
\label{gradeqe}
&& \frac{\dd \omega}{\dd q}=-(\grad V)^\omega\simeq -\omega^4 \pd_\omega V~~\nn\\
&& \frac{\dd \theta}{\dd q}=-(\grad V)^\theta\simeq -\frac{1}{{\tilde c}_\e}e^{-\frac{2\epsilon_\e}{\omega}}\pd_\theta V~~.
\eeqan
We studied these equations in \cite{grad} for all ends of
$\Sigma$. Below, we summarize the results for critical ends (see
op. cit. for the noncritical ends).

Recall that $V$ is globally well-behaved and $\hV$ is Morse on
$\hSigma$. Together with the formulas above, this implies that
$(\grad_G V)^\omega$ tends to zero at all ends while $(\grad_G
V)^\theta$ tends to zero exponentially at flaring (i.e. non-cusp) ends
and to infinity at cusp ends. On the other hand, we have:
\be
\label{nue}
||\grad_G V||^2=||\dd V||^2=\frac{1}{\omega^4} (\partial_\omega V)^2+f_\e(1/\omega) (\partial_\theta V)^2\approx \frac{1}{\omega^4} (\partial_\omega V)^2+ {\tilde c}_\e e^{\frac{2\epsilon_\e}{\omega}} (\partial_\theta V)^2~~.
\ee
Thus $||\grad_G V||$ tends to infinity at all ends.

\subsection{Principal canonical coordinates centered at an end $\e$}

\begin{definition}
A canonical Cartesian coordinate system $(x,y)$ for $(\Sigma,G)$
centered at the critical end $\e$ is called {\em principal} for $V$ if
the tangent vectors $\epsilon_x=\frac{\pd}{\pd x}\bvert_\e$ and
$\epsilon_y=\frac{\pd}{\pd y}\bvert_\e$ form a principal basis for $V$
at $\e$.
\end{definition}

\noindent Canonical Cartesian coordinates $(x,y)$ centered at $\e$ are given by:
\be
x=\omega\cos \theta=\frac{1}{r}\cos \theta~~~ ,~~~ y=\omega\sin\theta=\frac{1}{r}\sin\theta~.
\ee
In such coordinates, the end $\e$ corresponds to $\omega=0$,
i.e. $(x,y)=(0,0)$. The Taylor expansion of  $\hV$ at $\e$ in principal Cartesian coordinates $(x,y)$ centered at $\e$ and in  associated polar coordinates $(\omega,\theta)$ reads:
\beqan
\label{Vase}
\hV_\e(x,y)\!&=&\hV(\e)+\frac{1}{2}\!\left[\lambda_1(\e) x^2\!+\!\lambda_2(\e) y^2\right]+\O((x^2+y^2)^\frac{3}{2})~,\nn\\
\!\!\hV_\e(\omega,\theta)\!&=&\!\hV(\e)+\frac{1}{2}\omega^2\!\left[\lambda_1(\e) \cos^2 \theta\!+\!\lambda_2(\e)\sin^2 \theta\right]+\O(\omega^3)~,
\eeqan
where $\omega=\sqrt{x^2+y^2}$, $\theta= \arg(x+\i y)$ and the real numbers $\lambda_1(\e)$ and
$\lambda_2(\e)$ are the
principal values of the Hessian of $\hV(\e)$. When $\lambda_1$ and $\lambda_2$ do not
both vanish, it is convenient to define:

\begin{definition}
The {\em critical modulus} of $(\Sigma,\!G,\!V)$ at the
critical end $\e$ is the ratio:
\be
\label{beta}
\beta_\e\eqdef \frac{\lambda_1(\e)}{\lambda_2(\e)}\in [-1,1]\setminus \{0\}~~,
\ee
where $\lambda_1(\e)$ and $\lambda_2(\e)$ are the principal values of
$(\Sigma,G,V)$ at $\e$. 
\end{definition}
\begin{definition}
The {\em characteristic signs} of $(\Sigma,G,V)$ at $\e$ are:
\be
\varepsilon_i(\e)\eqdef \sign(\lambda_i(\e))\in \{-1,1\}~~(i=1,2)~~.
\ee
\end{definition}

The extended scalar potential $\hPhi$ of the canonical model can be
recovered from the extended classical effective potential as:
\ben
\label{Phic}
\hPhi=\frac{1}{2 M_0^2} \hV^2\approx \frac{{\bar \lambda}_2(\e)^2}{2}\left[\hat{\bar{V}}(\e)+\frac{1}{2}\omega^2 (\beta_\e \cos^2\theta+\sin^2\theta)\right]^2~,
\een
where we defined 
\be
{\bar \lambda}_2(\e)\eqdef \frac{\lambda_2(\e)}{M_0}~~,~~{\hat {\bar V}}(\e)\eqdef \frac{{\hat V}(\e)}{\lambda_2(\e)}~~.
\ee
Solving the gradient flow equation \eqref{gradV} of $V$ relative to
$G$ with the approximations \eqref{feas} and \eqref{Vase} for
$\theta\not \in \{0,\frac{\pi}{2},\pi,\frac{3\pi}{2}\}$ shows that the
{\em unoriented gradient flow orbits} of $V$ around the end $\e$ have
implicit equation:
\ben
\label{gammaflow}
\frac{1}{4}[\lambda_1(\e)-\lambda_2(\e)]\,\Gamma_2\!
\left(\frac{2\varepsilon_\e}{\omega}\right)=A+{\tilde c}_\e\left[\,\lambda_1(\e)\log|\sin\theta|-\lambda_2(\e)\log|\cos\theta |\,\right]~~,
\een
where $\Gamma_2$ is the lower incomplete Gamma function of order $2$ and $A$ is an integration constant.

Below, we compare graphically (making certain choices for $\beta$) the unoriented gradient flow orbits given implicitely by equation \eqref{gammaflow} to the orbits of {\em IR optimal cosmological curves}, defined as those
solutions $\varphi$ of the cosmological equation \eqref{cosm} which satisfy
$\dot\varphi(0)=-(\grad_G V)(\varphi(0))$. 

\subsection{The IR behavior near critical plane ends}

\begin{figure}[H]
\centering
\begin{minipage}{.47\textwidth}
\centering \includegraphics[width=.99\linewidth]{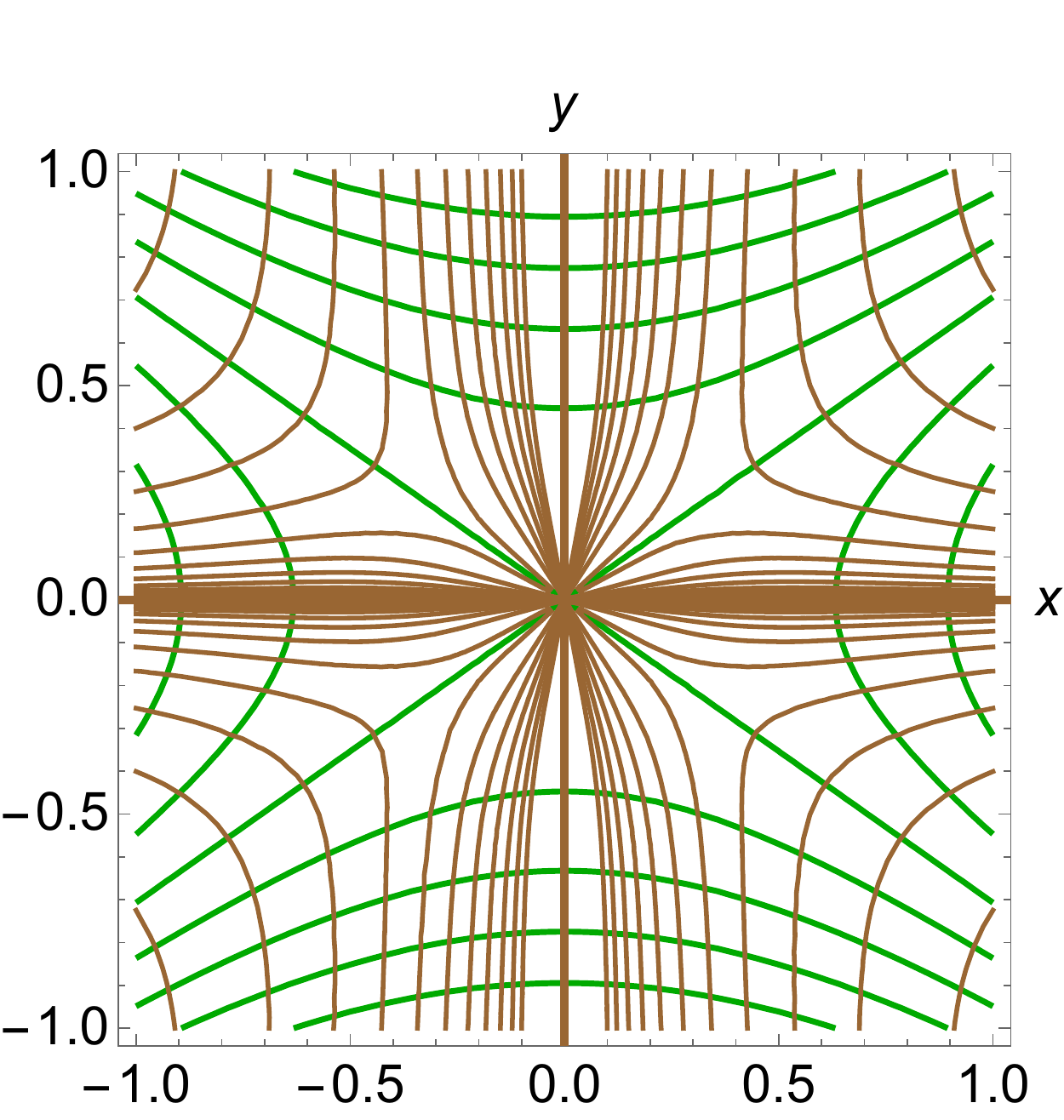}
\subcaption{For $\beta_\e=-0.5$}
\end{minipage}\hfill 
\begin{minipage}{.47\textwidth}
\centering \includegraphics[width=.99\linewidth]{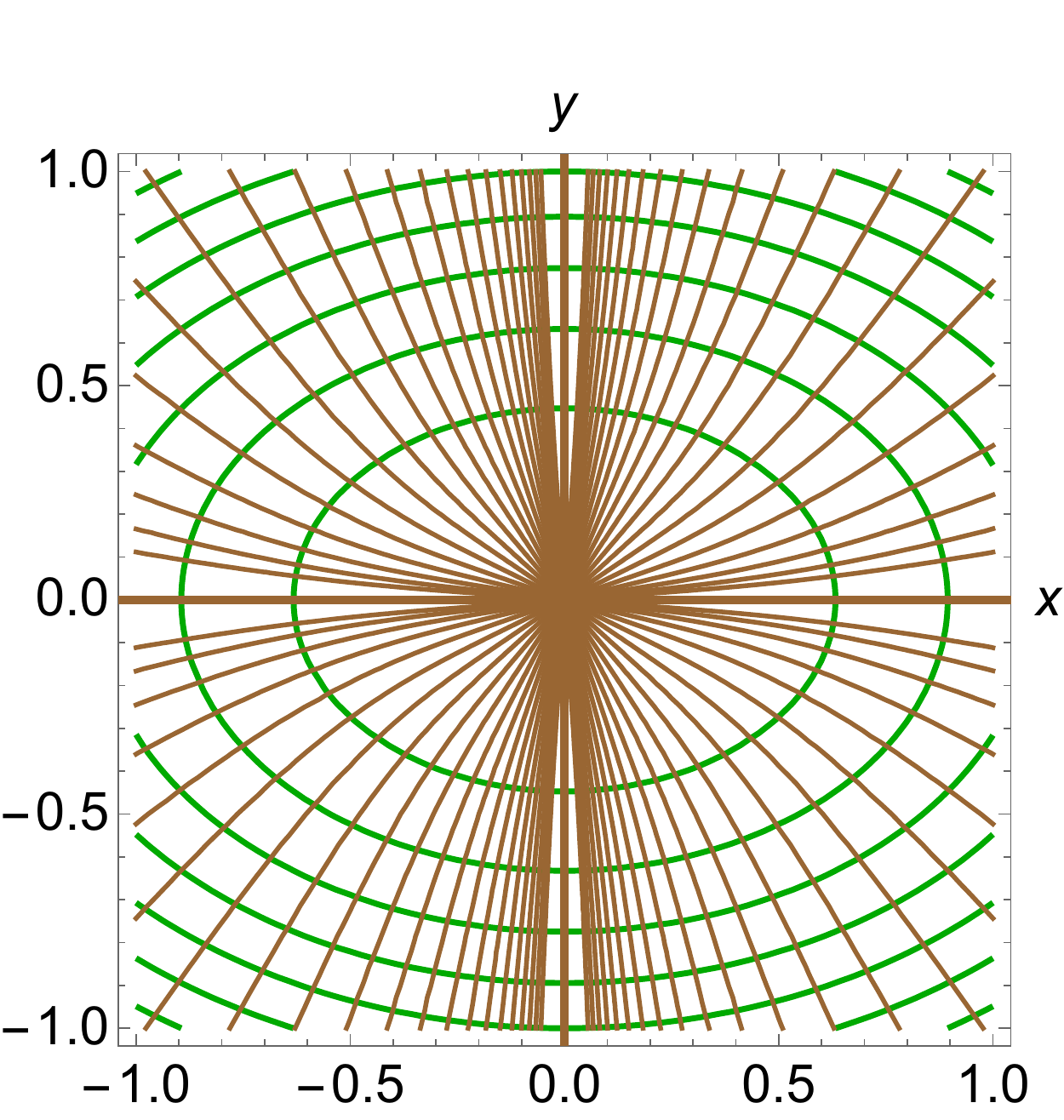}
\subcaption{For $\beta_\e=0.5$}
\end{minipage}
\caption{Gradient flow orbits of $V$ (shown in brown) and level sets
of $V$ (shown in green) near a critical plane end $\e$, drawn in
principal Cartesian canonical coordinates centered at $\e$ for two
values of $\beta_\e$.}
\label{fig:CritPlanePQ}
\end{figure}

\begin{figure}[H]
\begin{minipage}{.47\textwidth}
\centering  \includegraphics[width=.99\linewidth]{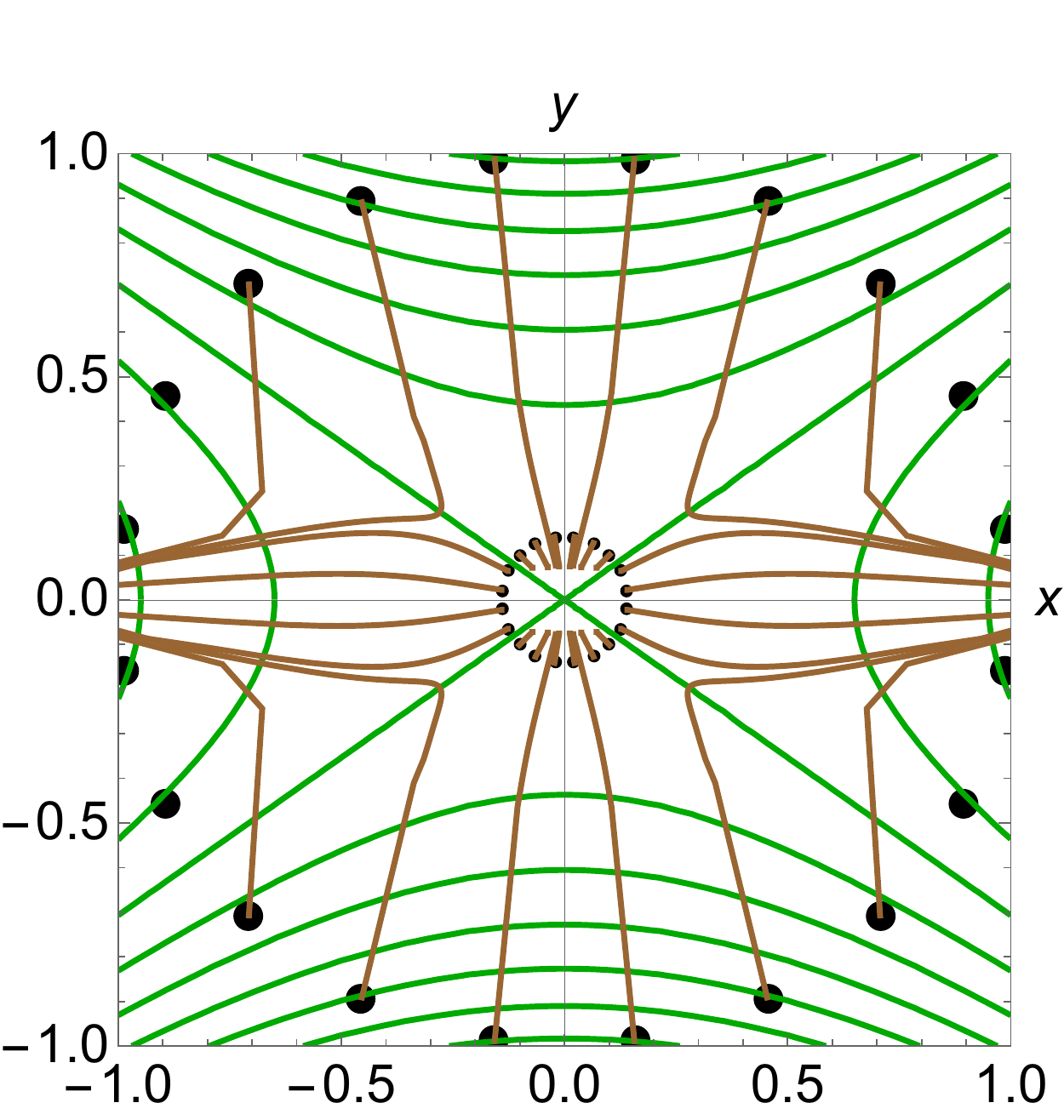}
\subcaption{For $\beta_\e=-0.5$}
\end{minipage}\hfill 
\begin{minipage}{.47\textwidth}
\vspace{2.5em}
\centering y\vspace{-1.5em} \includegraphics[width=.99\linewidth]{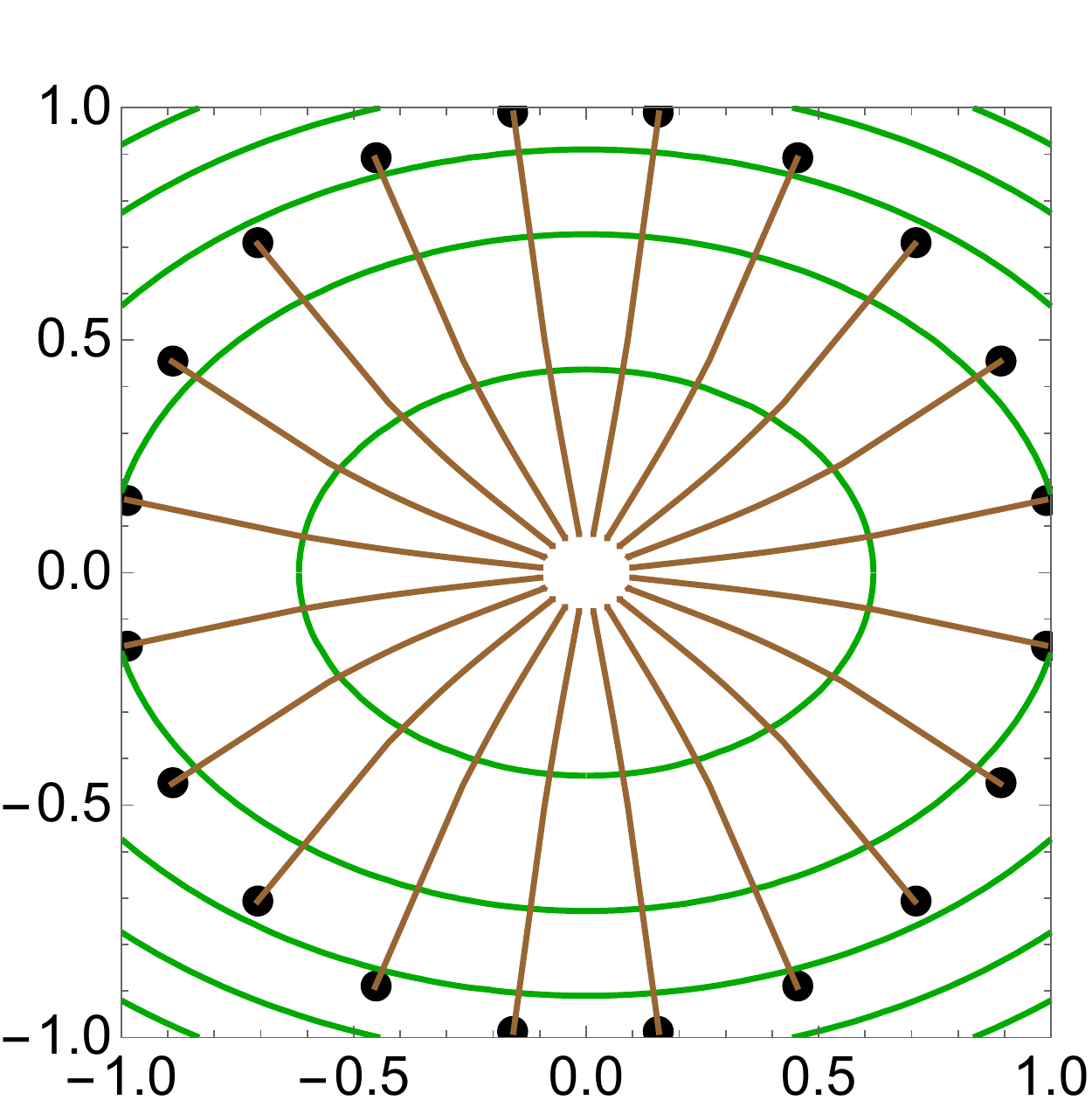}
\subcaption{For $\beta_\e=0.5$}
\end{minipage}
\caption{Numerically computed infrared optimal cosmological orbits of
the canonical model (shown in brown) and level sets of $\hPhi$ (shown
in green) near a critical plane end $\e$, drawn in principal canonical
Cartesian coordinates centered at $\e$ for two values of $\beta_\e$. We took ${\bar \lambda}_2(\e)=1$, ${\hat {\bar
V}}(\e)=1$ and $M_0=1$. The initial
point of each orbit, i.e. $\varphi(0)$, is shown as a black dot. }
\label{fig:CritCosmPlane}
\end{figure}

Figure \ref{fig:CritPlanePQ} bellow gives the unoriented gradient flow orbits for certain choices of $\beta$, while Figure \ref{fig:CritCosmPlane} gives the numerically computed orbits of the IR optimal cosmological curves for the same choices of $\beta$ and various other assumptions mentioned in the description. 

\vspace{-2.5em}

\subsection{The IR behavior near critical horn ends}

\begin{figure}[H]
\centering
\begin{minipage}{.46\textwidth}
\centering  \includegraphics[width=.99\linewidth]{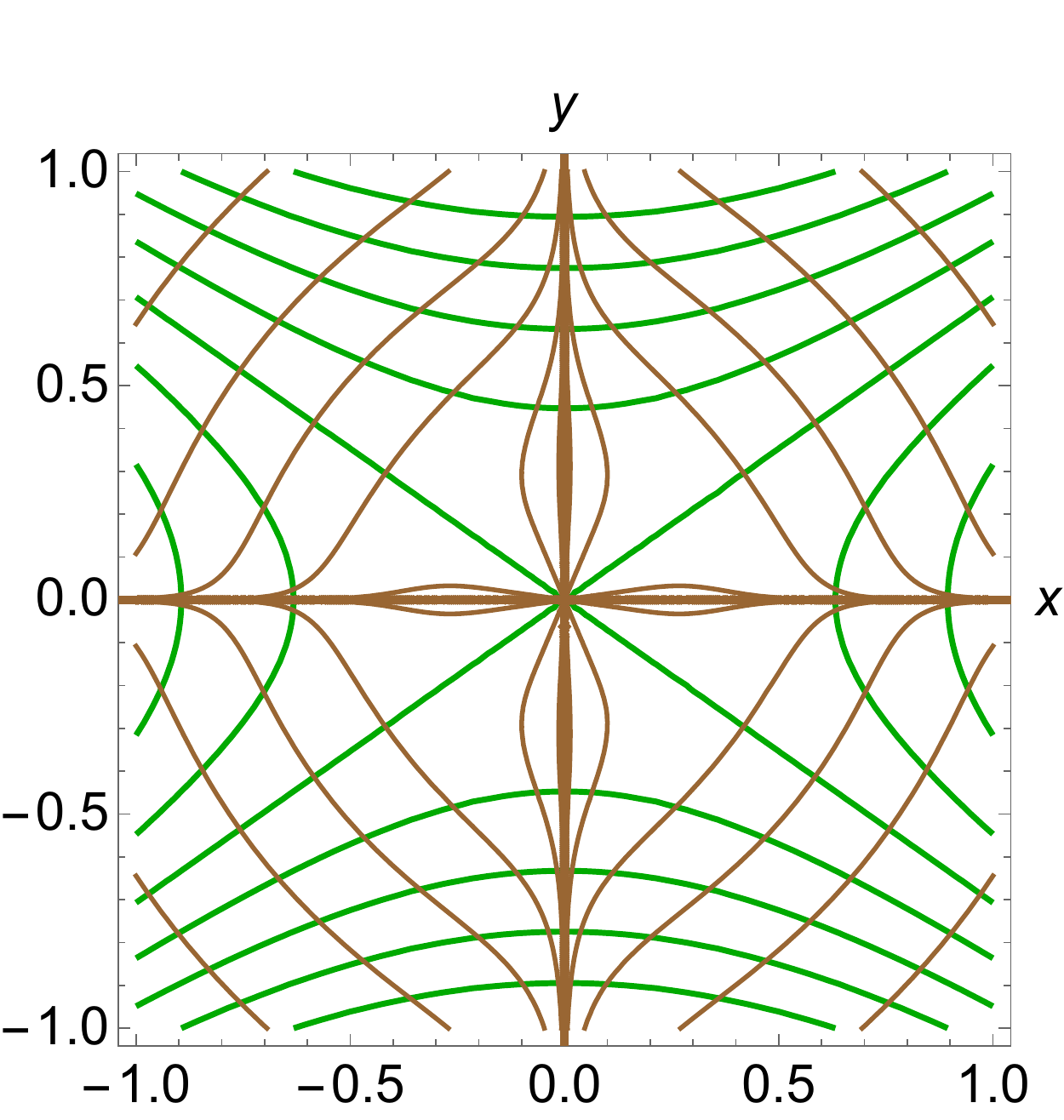}
\subcaption{For $\beta_\e=-0.5$.}
\end{minipage}\hfill
\begin{minipage}{.47\textwidth}
\centering \includegraphics[width=.99\linewidth]{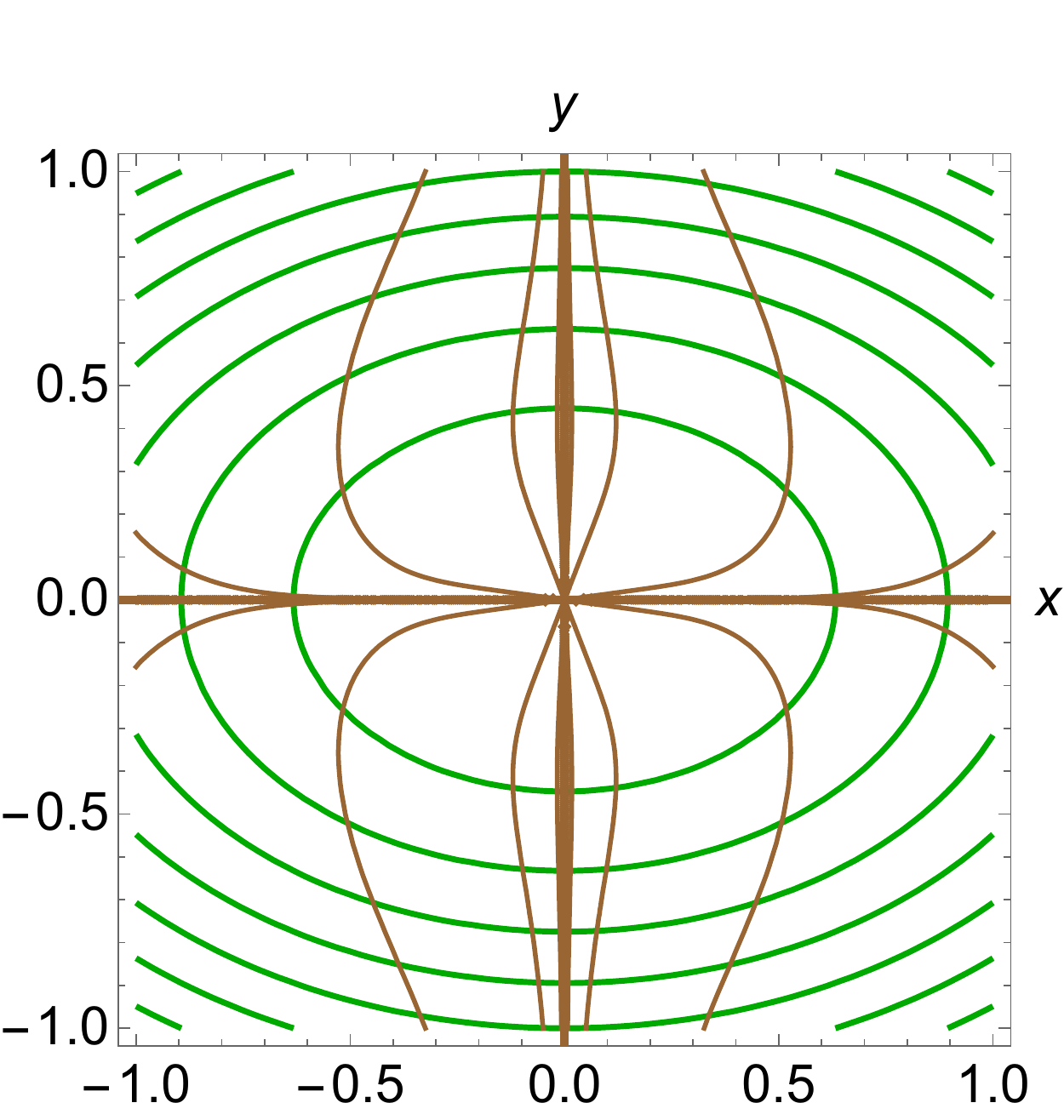}
\subcaption{For $\beta_\e=0.5$.}
\end{minipage}
\caption{Gradient flow orbits of $V$ (shown in brown) and level sets
of $V$ (shown in green) near a critical horn end $\e$, drawn in
principal Cartesian canonical coordinates centered at $\e$ for two
values of $\beta_\e$.}
\label{fig:CritHornPQ}
\end{figure}

\begin{figure}[H]
\centering
\begin{minipage}{.46\textwidth}
\centering  \includegraphics[width=.99\linewidth]{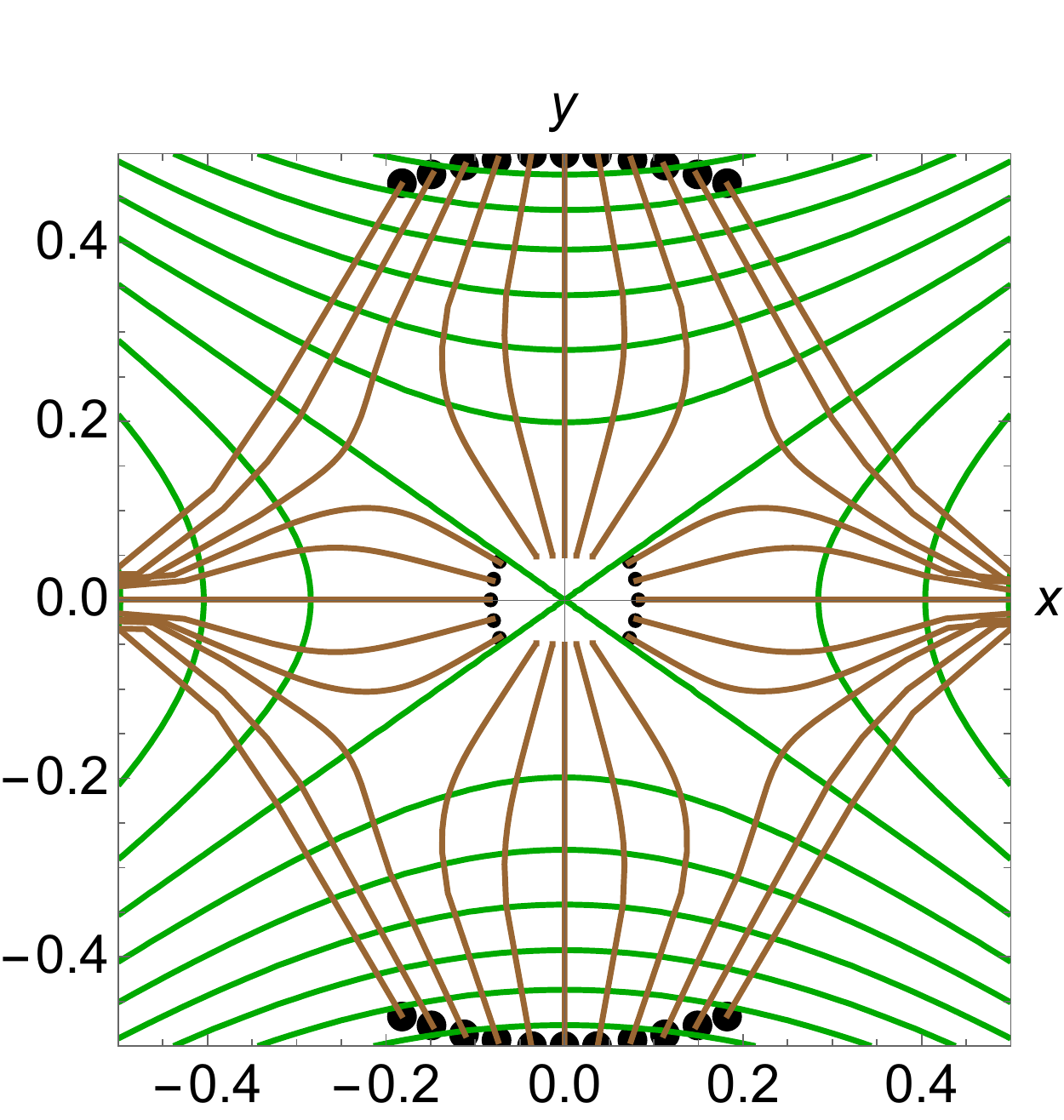}
\subcaption{For $\beta_\e=-0.5$.}
\end{minipage}\hfill
\begin{minipage}{.47\textwidth}
\centering \includegraphics[width=.99\linewidth]{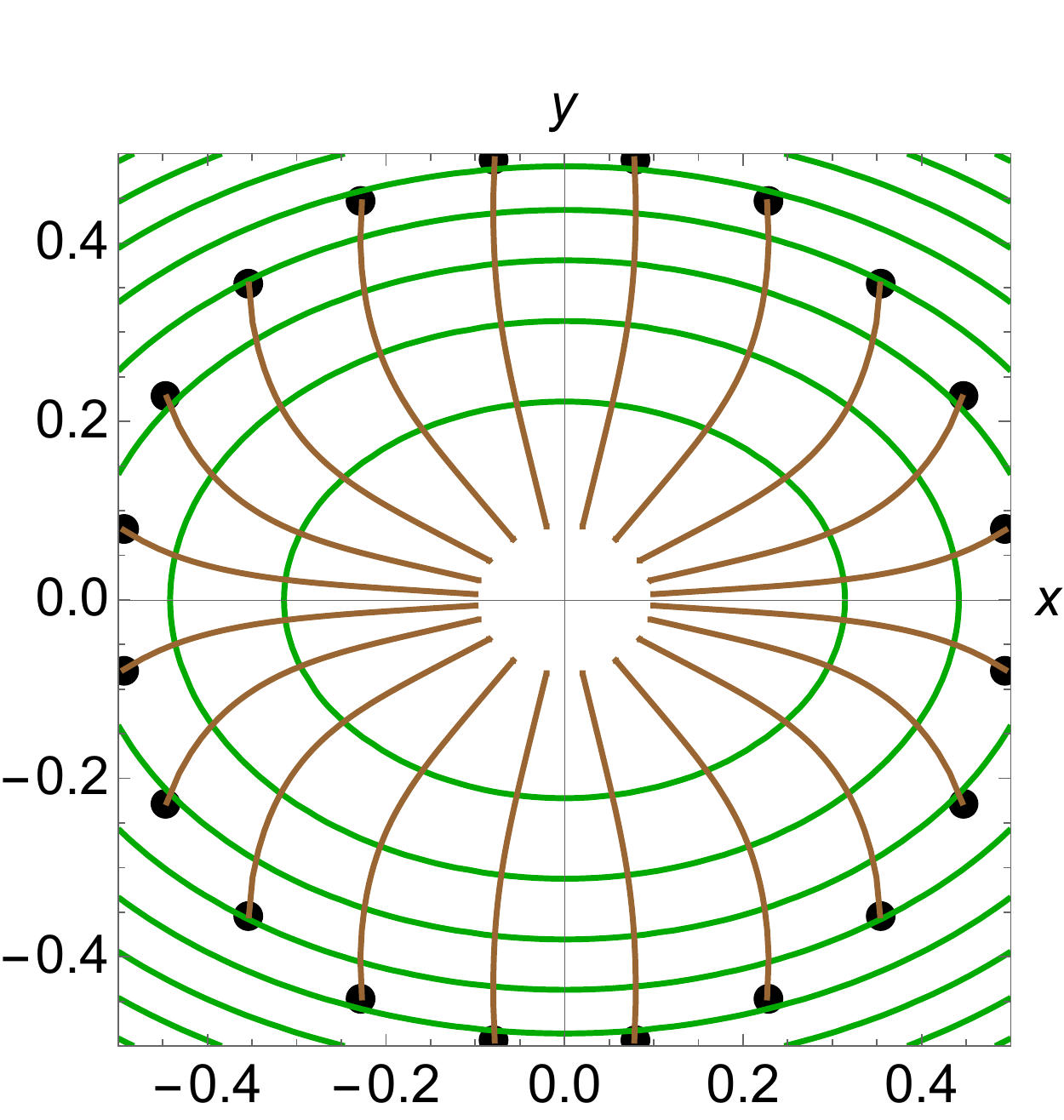}
\subcaption{For $\beta_\e=0.5$.}
\end{minipage}
\caption{Numerically computed infrared optimal cosmological orbits of
the canonical model (shown in brown) and level sets of $\hPhi$ (shown
in green) near a critical horn end $\e$, drawn in principal canonical
Cartesian coordinates centered at $\e$ for two values of $\beta_\e$. 
We took ${\bar \lambda}_2(\e)=1$, ${\hat {\bar V}}(\e)=1$ and $M_0=1$. 
The initial point $\varphi(0)$ of each orbit is shown as a black dot.}
\label{fig:CritCosmHorn}
\end{figure}

We graphically compare Figure \ref{fig:CritHornPQ} bellow, which  gives the unoriented gradient flow orbits near critical horn ends, with Figure \ref{fig:CritCosmHorn} which shows some numerically computed orbits of the IR optimal cosmological curves near critical horn ends. The comparison is done for certain choices of $\beta$. 

\vspace{-2em}

\subsection{The IR behavior near critical funnel ends}
  
\begin{figure}[H]
\centering
\vskip -1em
\begin{minipage}{.47\textwidth}
\centering  \includegraphics[width=.99\linewidth]{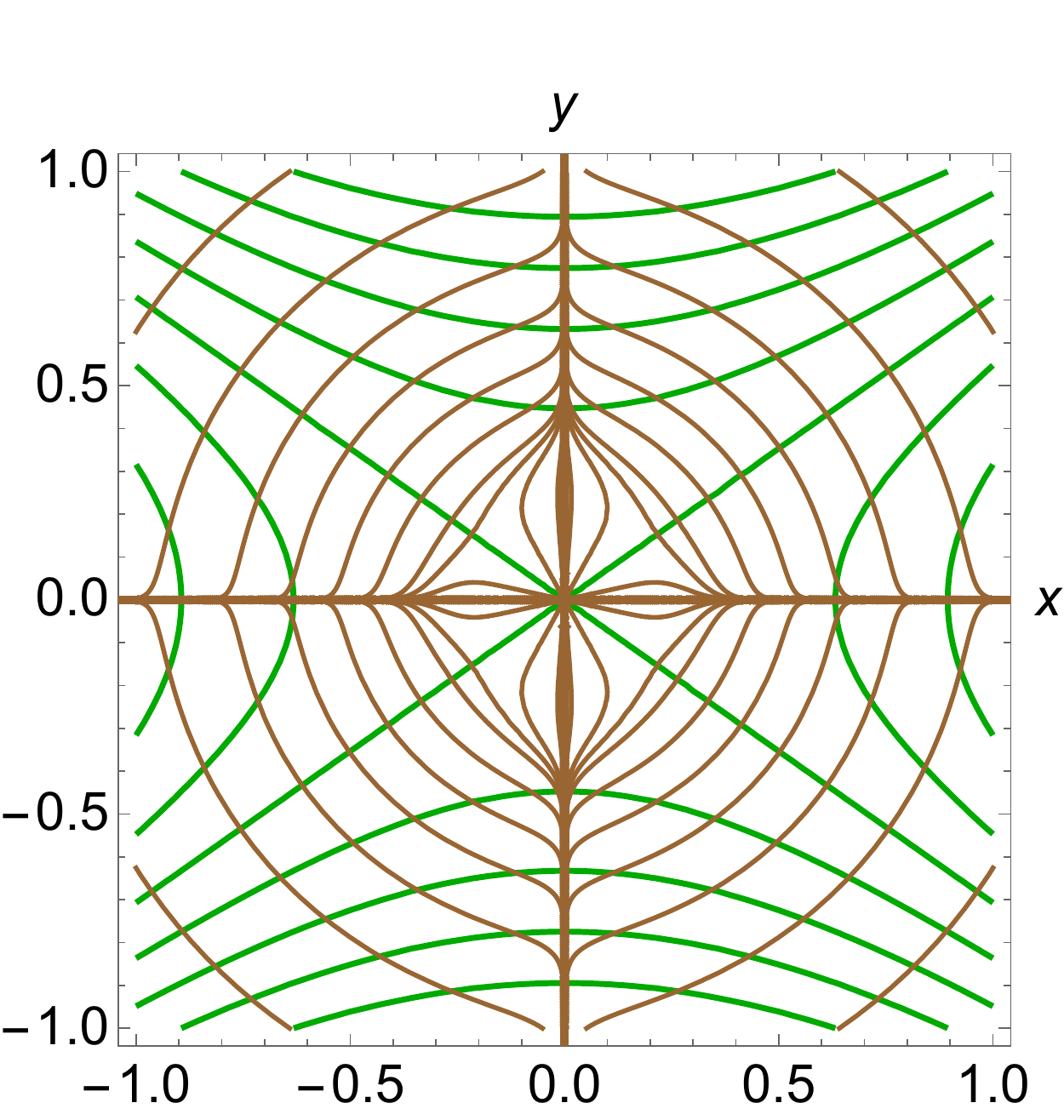}
\subcaption{For $\beta_\e=-0.5$.}
\end{minipage}\hfill
\begin{minipage}{.47\textwidth}
\centering \includegraphics[width=.99\linewidth]{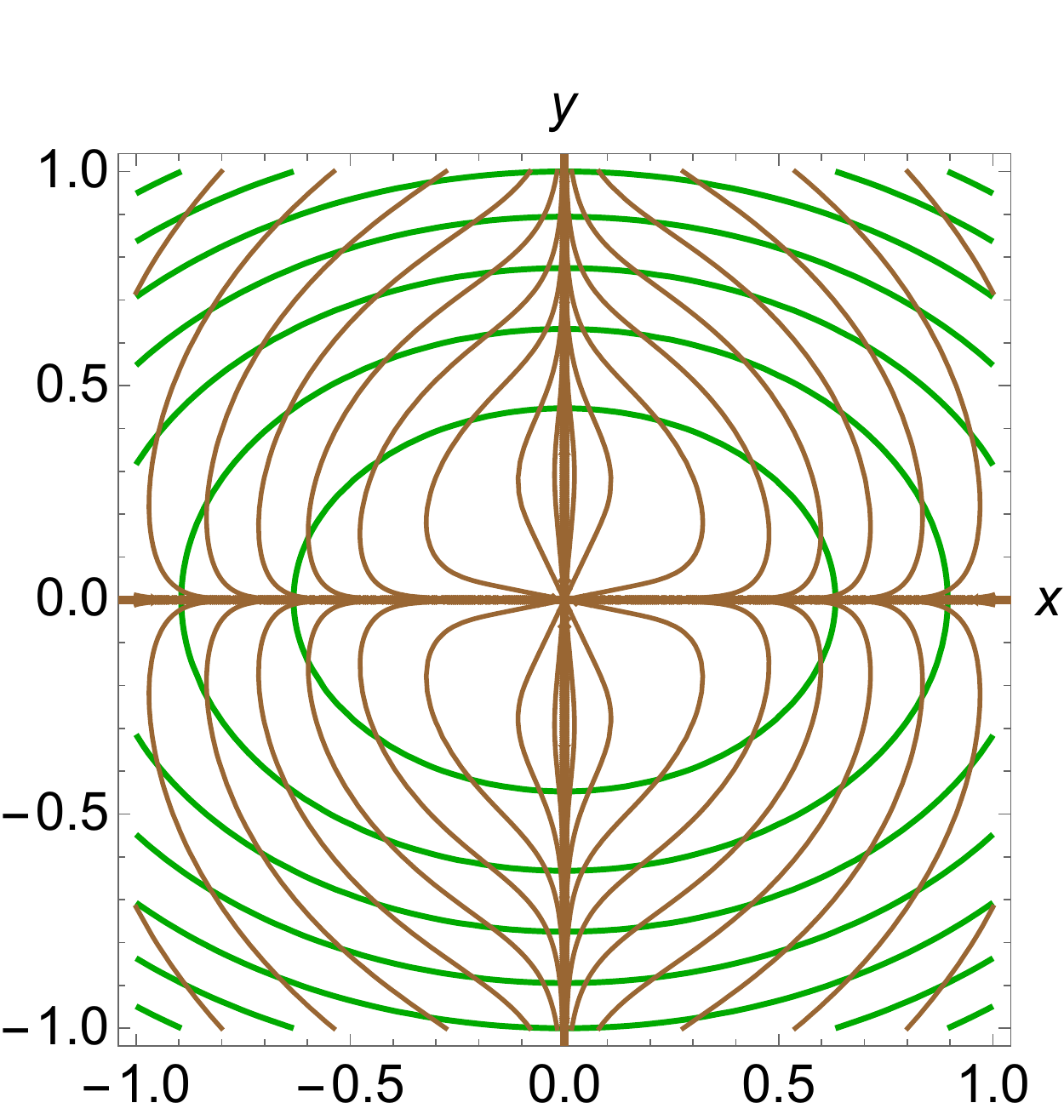}
\subcaption{For $\beta_\e=0.5$.}
\end{minipage}
\caption{Gradient flow orbits of $V$ (shown in brown) and level sets
of $V$ (shown in green) near a critical funnel end $\e$ of
circumference $\ell=1$, drawn in principal Cartesian canonical
coordinates centered at $\e$ for two values of $\beta_\e$. }
\label{fig:CritFunnelPQ}
\end{figure}

\begin{figure}[H]
\centering
\begin{minipage}{.47\textwidth}
\centering  \includegraphics[width=.99\linewidth]{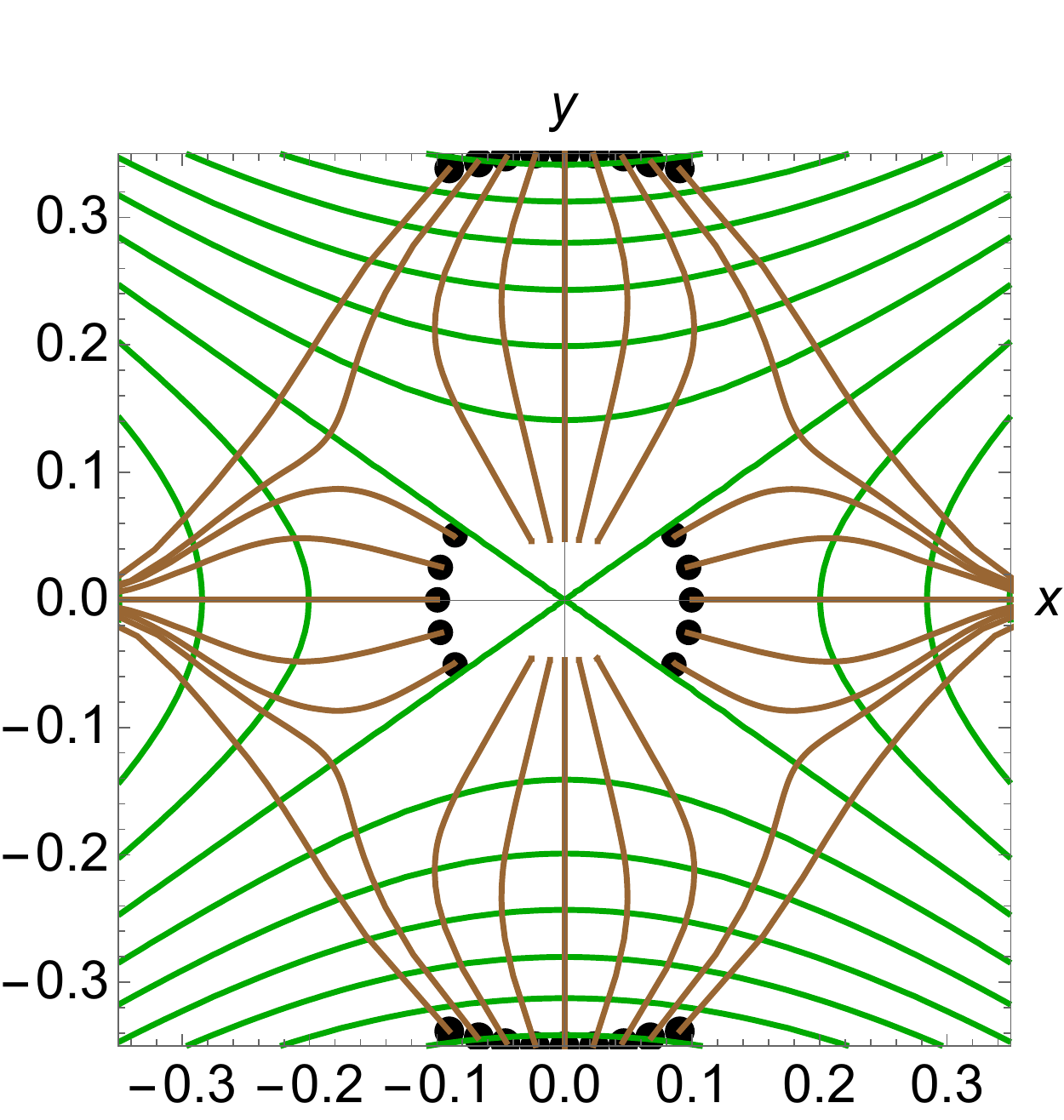}
\subcaption{For $\beta_\e=-0.5$.}
\end{minipage}\hfill
\begin{minipage}{.47\textwidth}
\centering \includegraphics[width=.99\linewidth]{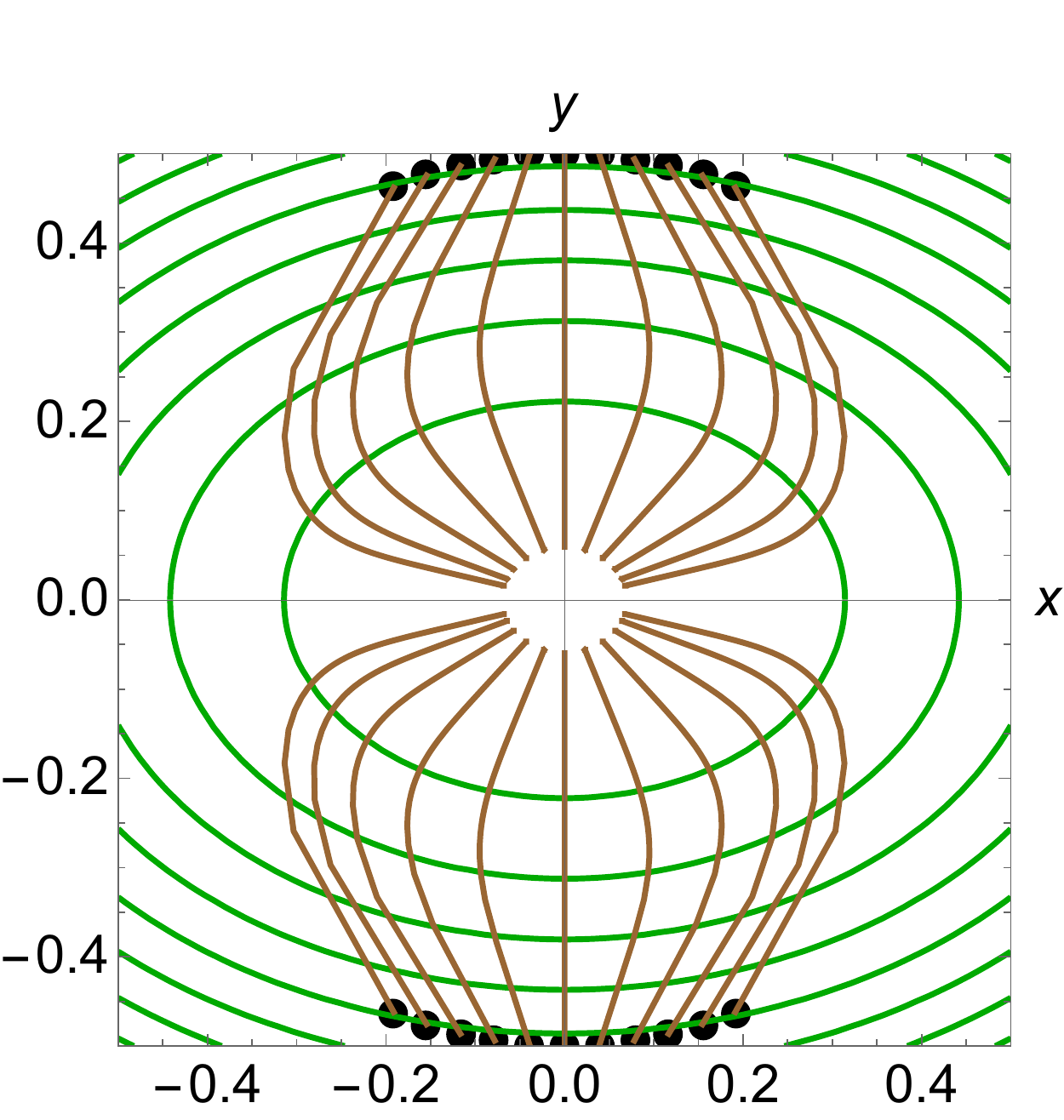}
\subcaption{For $\beta_\e=0.5$.}
\end{minipage}
\caption{Numerically computed infrared optimal cosmological orbits of the
canonical model (shown in brown) and level sets
of $\hPhi$ (shown in green) near a critical funnel end $\e$ of
circumference $\ell=1$, drawn in principal canonical coordinates
centered at $\e$ for two values of $\beta_\e$. We took ${\bar \lambda}_2(\e)=1$, ${\hat {\bar
V}}(\e)=1$ and $M_0=1$. The initial point $\varphi(0)$ of each orbit is shown as a black dot.}
\label{fig:CritCosmFunnel}
\end{figure}

  We visually compare below Figure \ref{fig:CritFunnelPQ}, which  gives the unoriented gradient flow orbits near critical funnel ends, with Figure \ref{fig:CritCosmFunnel}, which shows some numerically computed orbits of the IR optimal cosmological curves near critical funnel ends. 
  
\vspace{-2em}

\subsection{The IR behavior near critical cusp ends}

\begin{figure}[H]
\centering

\begin{minipage}{.47\textwidth}
\centering  \includegraphics[width=.99\linewidth]{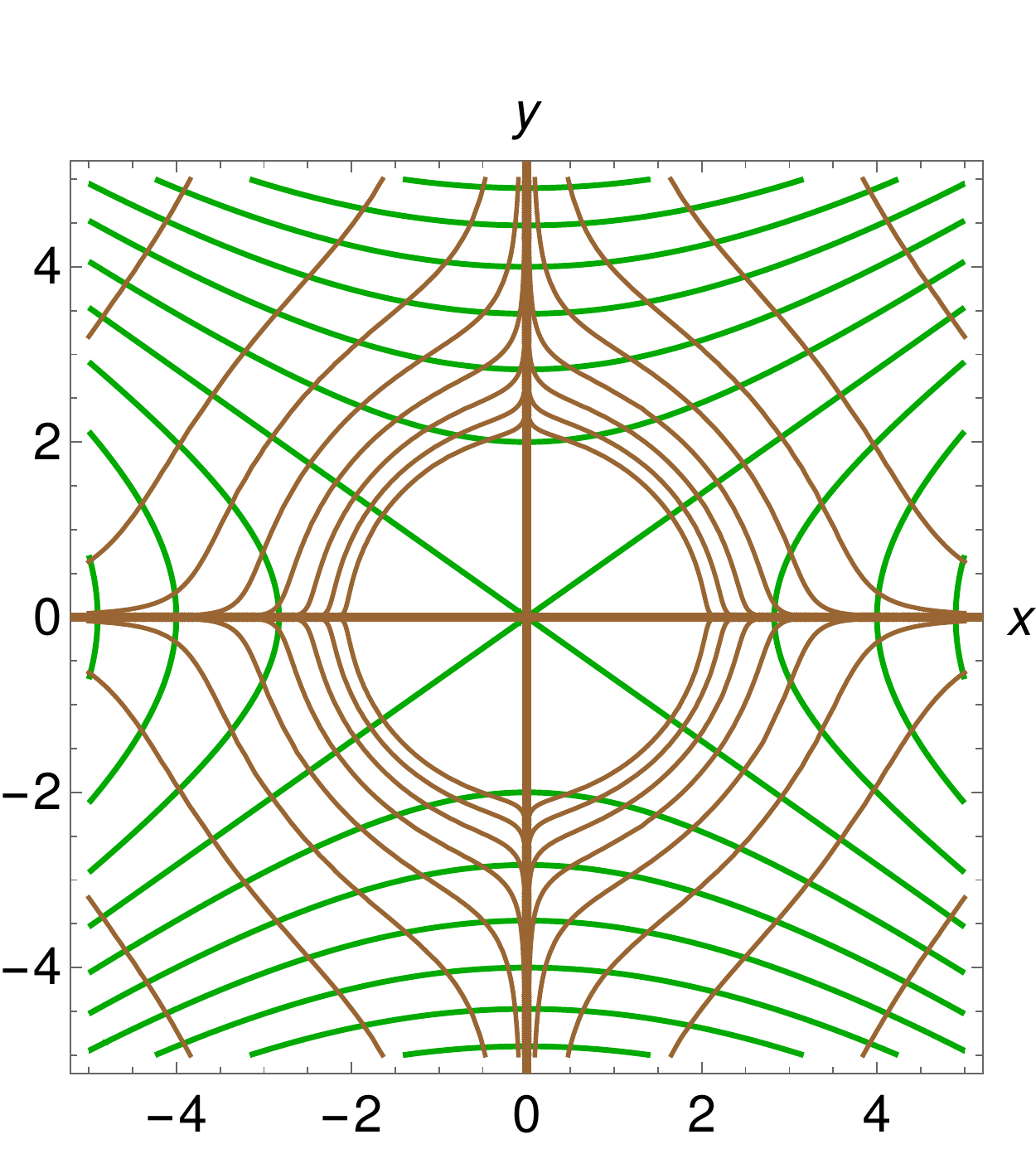}
\subcaption{For $\beta_\e=-0.5$.}
\end{minipage}\hfill
\begin{minipage}{.47\textwidth}
\centering \includegraphics[width=.99\linewidth]{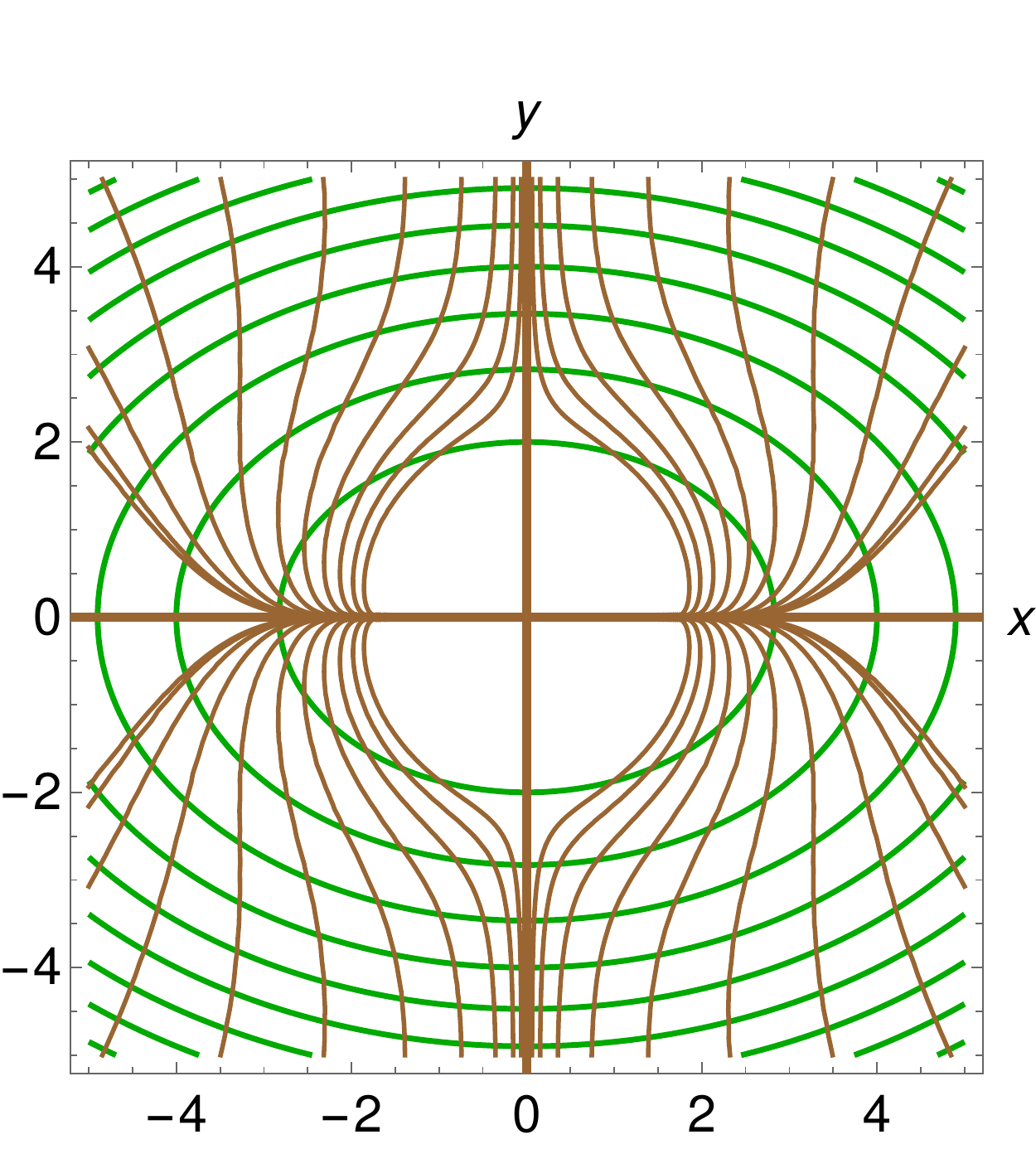}
\subcaption{For $\beta_\e=0.5$.}
\end{minipage}
\caption{Gradient flow orbits of $V$ (shown in brown) and level sets
of $V$ (shown in green) near a critical cusp end $\e$, drawn in
principal Cartesian canonical coordinates centered at $\e$ for two
values of $\beta_\e$.}
\label{fig:CritCuspPQ}
\end{figure}

\begin{figure}[H]
\centering  
\begin{minipage}{.47\textwidth}
\centering  \includegraphics[width=.99\linewidth]{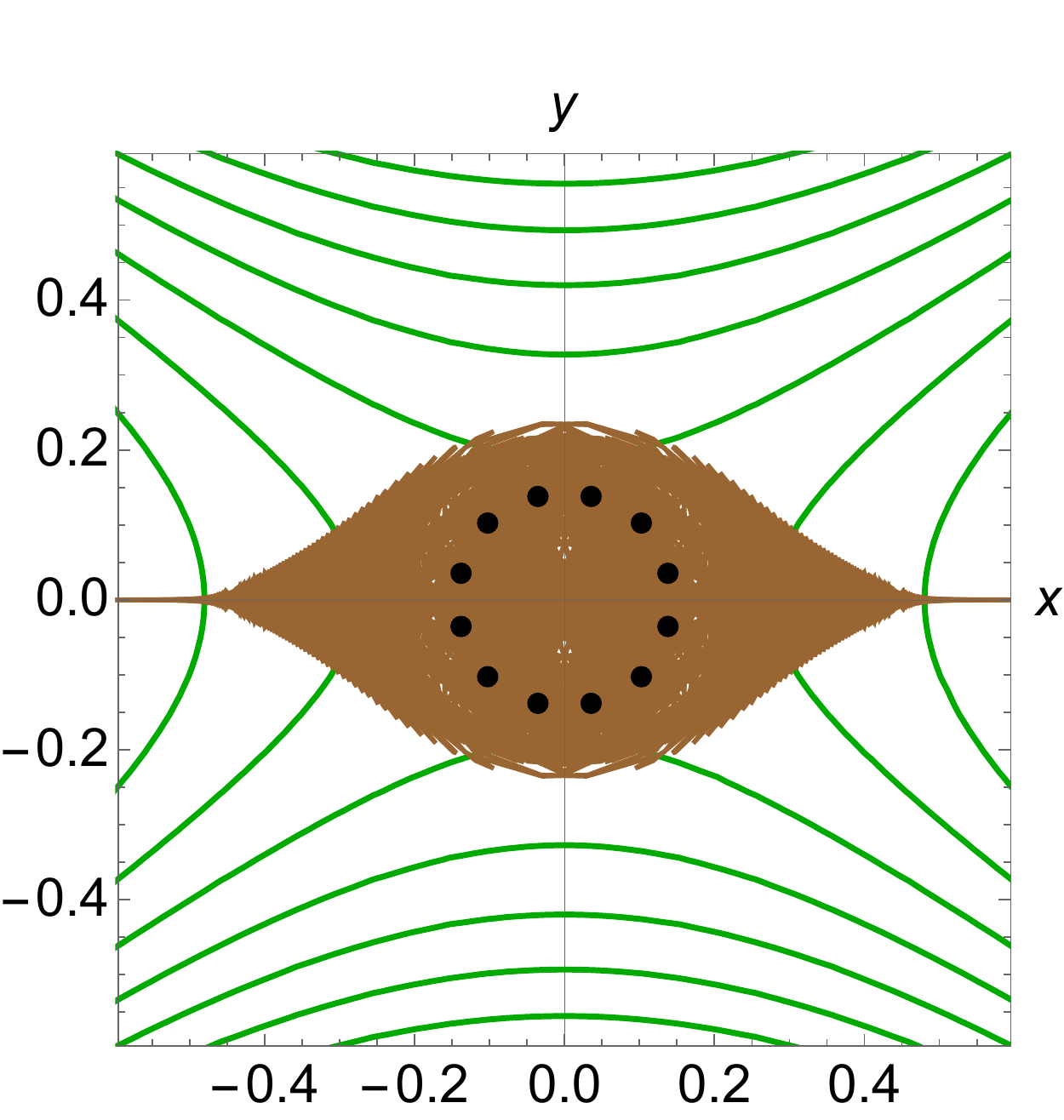}
\subcaption{For $\beta_\e=-0.5$.}
\end{minipage}\hfill
\begin{minipage}{.475\textwidth}
\centering \includegraphics[width=.99\linewidth]{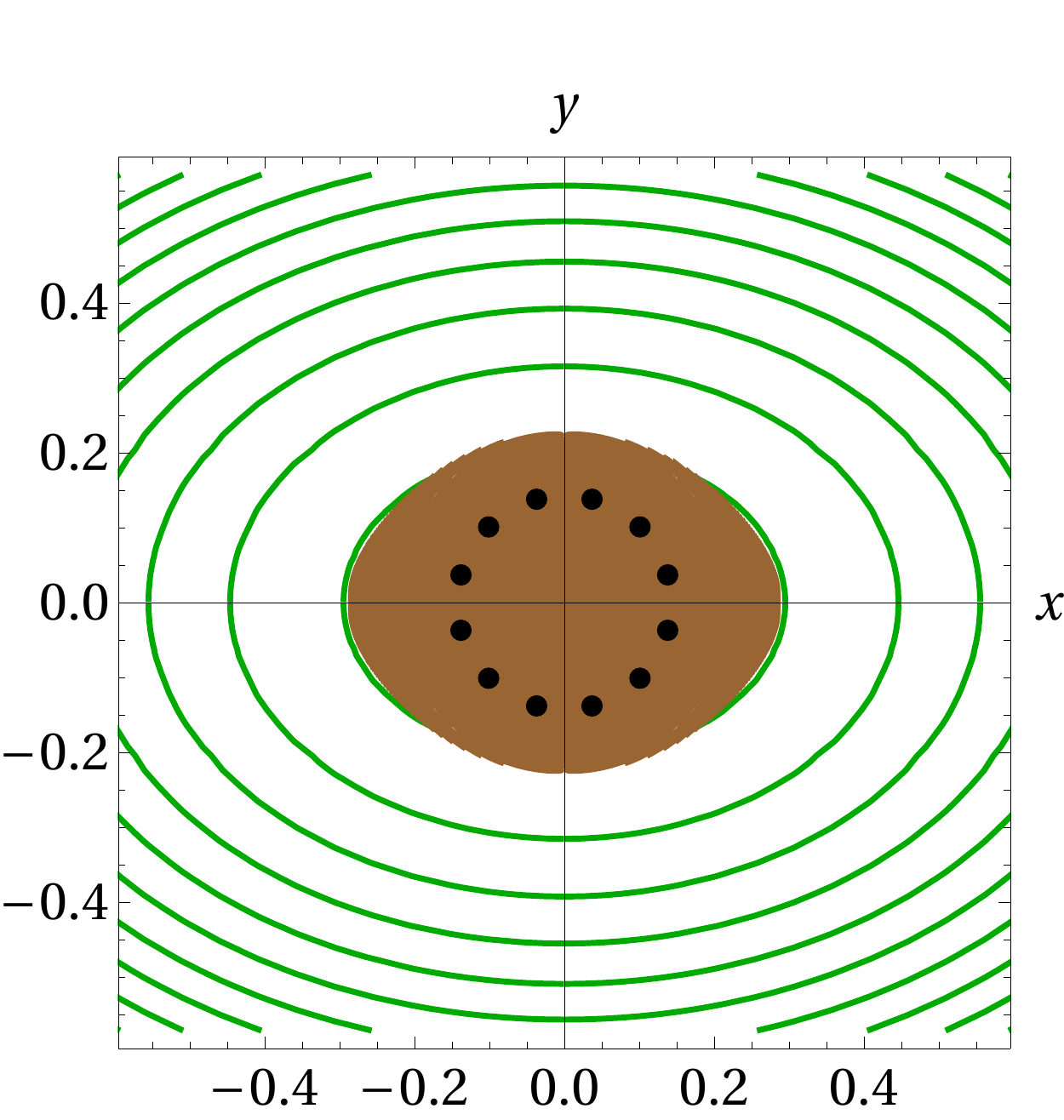}
\subcaption{For $\beta_\e=0.5$.}
\end{minipage}
\caption{Numerically computed infrared optimal cosmological orbits of the
canonical model (shown in brown) and level sets
of $\hPhi$ (shown in green) near a critical cusp end $\e$, drawn in
principal canonical coordinates centered at $\e$ for two values of
$\beta_\e$. We took ${\bar \lambda}_2(\e)=1$, ${\hat {\bar
V}}(\e)=1$ and $M_0=1$. The initial point $\varphi(0)$ of each orbit is shown as a black dot.}
\label{fig:CritCosmCusp}
\end{figure}

 By graphically comparing Figure \ref{fig:CritCuspPQ}, which  gives some unoriented gradient flow orbits near critical cusp ends, and Figure \ref{fig:CritCosmCusp}, which shows some numerically computed orbits of the IR optimal cosmological curves near critical cusp ends, one can assume that higher order corrections are needed in the IR expansion to get better approximants for the cusp.
  
  \vspace{-2em}

\vspace{-2em}

\subsection{The IR behavior near an interior critical point}

\noindent Let $\c$ be {\bf an interior critical point} and $(x,y)$ be
principal Cartesian canonical coordinates centered at $\c$. We have the metric:
\be
\dd s^2_G=\frac{4}{(1-\omega^2)^2}[\dd \omega^2+\omega^2\dd \theta^2]~~
\ee
and:
\be
V(\omega,\theta)= V(\c)+\frac{1}{2}\omega^2\left[\lambda_1(\c) \cos^2\theta +\lambda_2(\c) \sin^2\theta\right]+\O(\omega^3)~~,
\ee
where $\omega\eqdef\sqrt{x^2+y^2}$ and $\theta\eqdef \arg(x+\i y)$. Thus:
\beqan
\label{gradc}
&& (\grad V)^\omega\!\approx\!\frac{(1-\omega^2)^2}{4} \pd_\omega V\!=\!\frac{(1-\omega^2)^2\omega}{4}[\lambda_1(\c)\cos^2\theta+\lambda_2(\c)\sin^2\theta]~,\nn\\
&& (\grad V)^\theta\!\approx\!\frac{(1-\omega^2)^2}{4\omega^2} \pd_\theta V\!=\!\frac{(1-\omega^2)^2}{4} [\lambda_2(\c)\!-\!\lambda_1(\c)]\sin(\theta)\cos(\theta)~.
\eeqan
The {\em critical
modulus} $\beta_\c$ and {\em characteristic signs} $\epsilon_1(\c)$
and $\epsilon_2(\c)$ of $(\Sigma,G,V)$ at $\c$ are defined through:
\be
\beta_\c\eqdef\frac{\lambda_1(\c)}{\lambda_2(\c)}\in [-1,1]\setminus \{0\}~~,~~\epsilon_i(\c)\eqdef \sign(\lambda_i(\c))~~(i=1,2)~~.
\ee
Distinguish the cases:
\begin{enumerate}
\item $\lambda_1(\c)=\lambda_2(\c):=\lambda(\c)$,
  i.e. $\beta_\c=1$. Then $\epsilon_1(\c)=\epsilon_2(\c):=\epsilon(\c)$
and $\c$ is a local minimum of $V$ when $\lambda(\c)$ is positive
(i.e. when $\epsilon(\c)=1$) and a local maximum of $V$ when
$\lambda(\c)$ is negative (i.e. when $\epsilon(\c)=-1$). Relations
\eqref{gradc} become:
\be
(\grad V)^\omega\approx \frac{(1-\omega^2)^2\omega}{4}\lambda(\c)~~,~~ (\grad V)^\theta\approx 0
\ee
and the gradient flow equation of $(\Sigma,G,V)$ takes the following approximate form near $\c$:
\be
\frac{\dd \omega}{\dd q}=-\frac{(1-\omega^2)^2\omega}{4}\lambda(\c)~~,~~\frac{\dd \theta}{\dd q}=0~~.
\ee
This gives $\theta=\const$, i.e. the gradient flow curves near $\c$
are approximated by straight lines through the origin when drawn in
principal Cartesian canonical coordinates $(x,y)$ at $\c$; 
\item $\lambda_1(\c)\neq \lambda_2(\c)$, i.e. $\beta_\c\neq 1$.  When
$\theta\in \{0,\frac{\pi}{2}, \pi, \frac{3\pi}{2}\}$, the gradient
flow equation reduces to:
\beqa
&&\frac{\dd \omega}{\dd q}=\frac{(1-\omega^2)^2\omega}{4}\times \twopartdef{\lambda_1(\c)}{\theta\in \{0,\pi\}}{\lambda_2(\c)}{\theta\in \{\frac{\pi}{2},\frac{3\pi}{2}\}}\\
&& \frac{\dd \theta}{\dd q}=0~~.
\eeqa
This gives four gradient flow orbits which are approximated near $\c$
by the principal geodesic orbits. When $\theta\not \in
\{0,\frac{\pi}{2}, \pi, \frac{3\pi}{2}\}$, the gradient flow equation
takes the form:
\ben
\label{gradcn}
(1-\beta_\c)\frac{\dd \omega}{\dd \theta}=\omega (\beta_\c\cot\theta+\tan\theta)~~,
\een
with general solution:
\ben
\label{gradflowc}
\omega=C \frac{|\sin(\theta)|^{\frac{\beta_\c}{1-\beta_\c}}}{|\cos(\theta)|^\frac{1}{1-\beta_\c}}~~, ~~\theta\not \in
\{0,\frac{\pi}{2}, \pi, \frac{3\pi}{2}\}~~,~~C>0~.
\een
\end{enumerate}

\begin{figure}[H]
\centering
\begin{minipage}{.43\textwidth}
\centering  \includegraphics[width=0.99\linewidth]{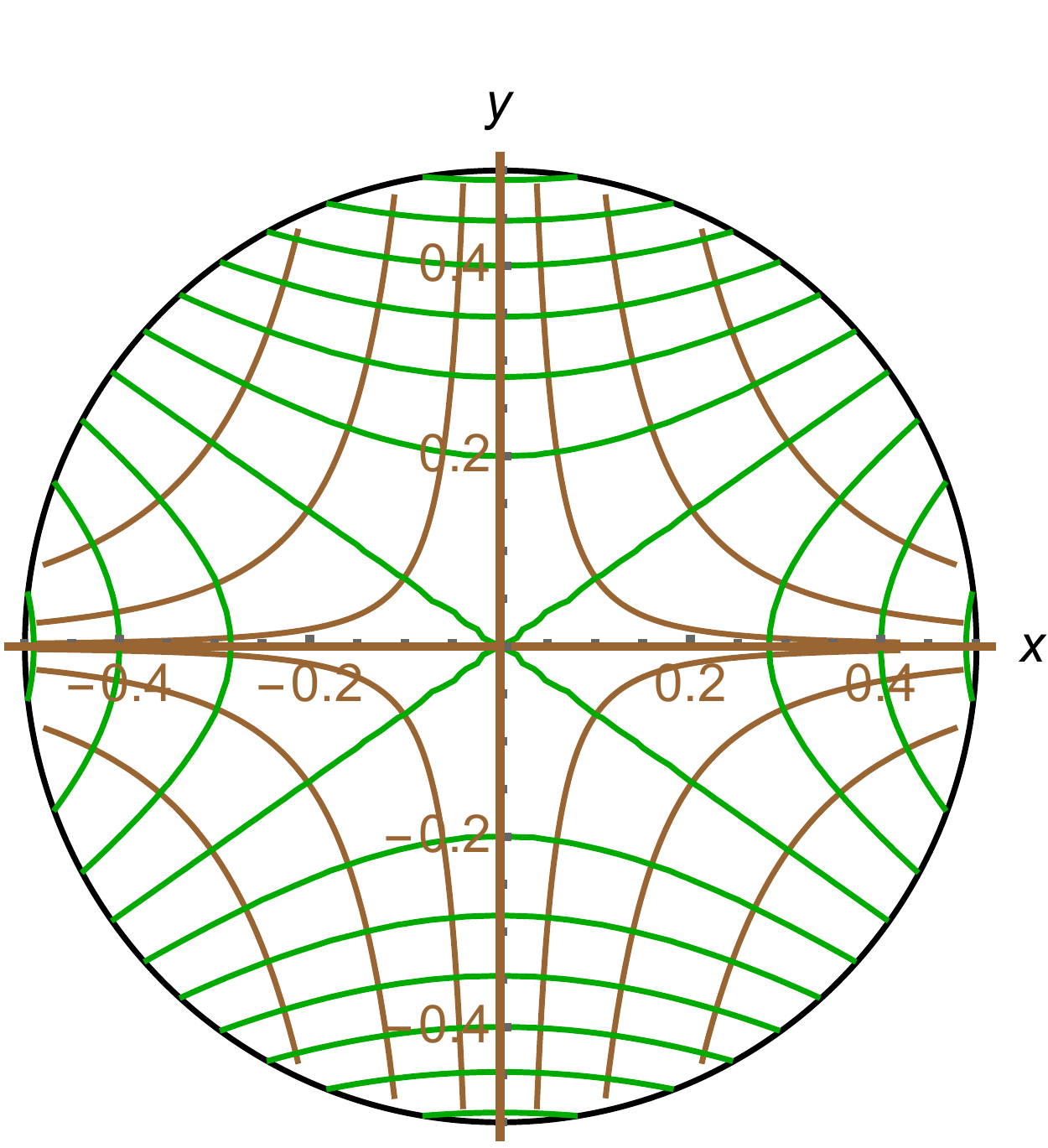}
\subcaption{$\beta_\c\!=\!-0.5$. Interior saddle point of $V$.}
\end{minipage}\hfill 
\begin{minipage}{.43\textwidth}
\centering \includegraphics[width=.99\linewidth]{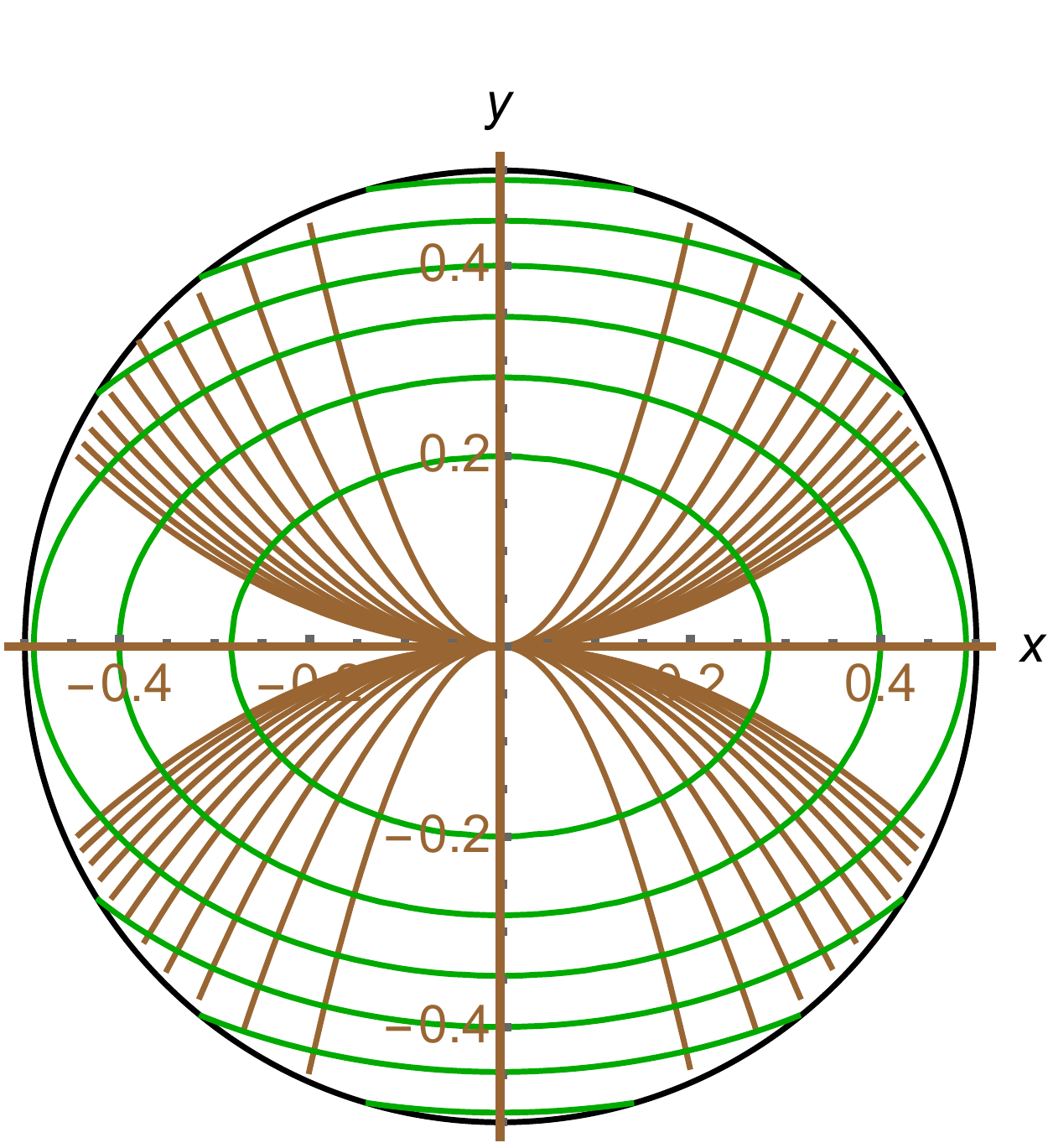}
\subcaption{$\beta_\c\!\!=\!0.5$. Local extremum of $V$.}
\end{minipage}
\caption{Unoriented gradient flow orbits of $V$ (shown in brown) near
an interior critical point superposed over the level lines of $V$
(shown in green) for two values of $\beta_\c$, plotted in principal
Cartesian canonical coordinates centered at the critical point. The figure
assumes $\omega_\rmax(\c)=0.5$. The principal coordinate
axes correspond to the principal geodesic orbits at $\c$, which
coincide with four special gradient flow orbits.}
\label{fig:IntCritical}
\end{figure}

\begin{figure}[H] \centering
\begin{minipage}{.43\textwidth} \centering
\includegraphics[width=.99\linewidth]{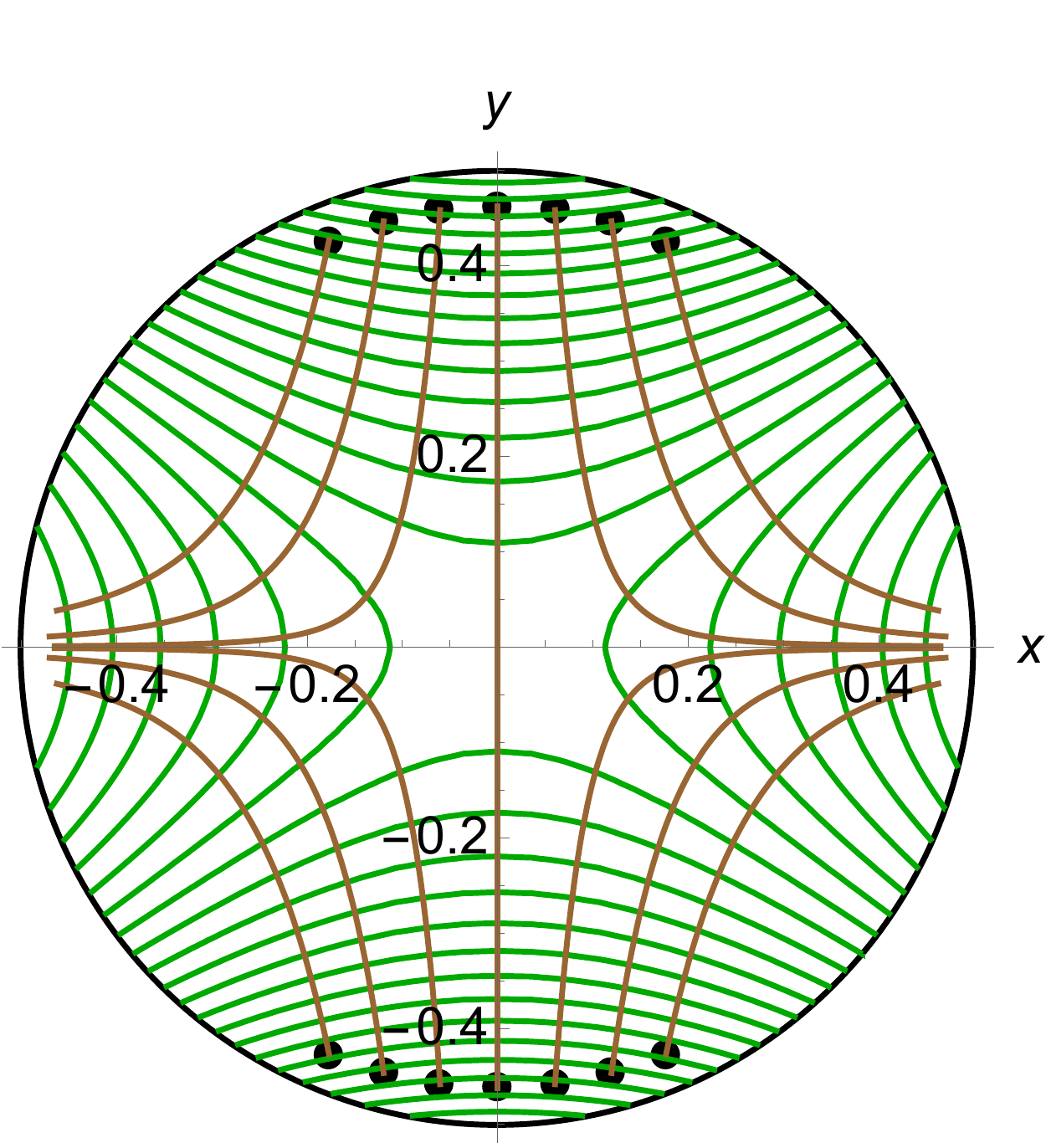} \subcaption{
Saddle point of $V$ for $\beta_\c\!=\!-0.5$. The dots are initial points $\varphi(0)$.}
\end{minipage}\hfill
\begin{minipage}{.43\textwidth} \centering
\includegraphics[width=.99\linewidth]{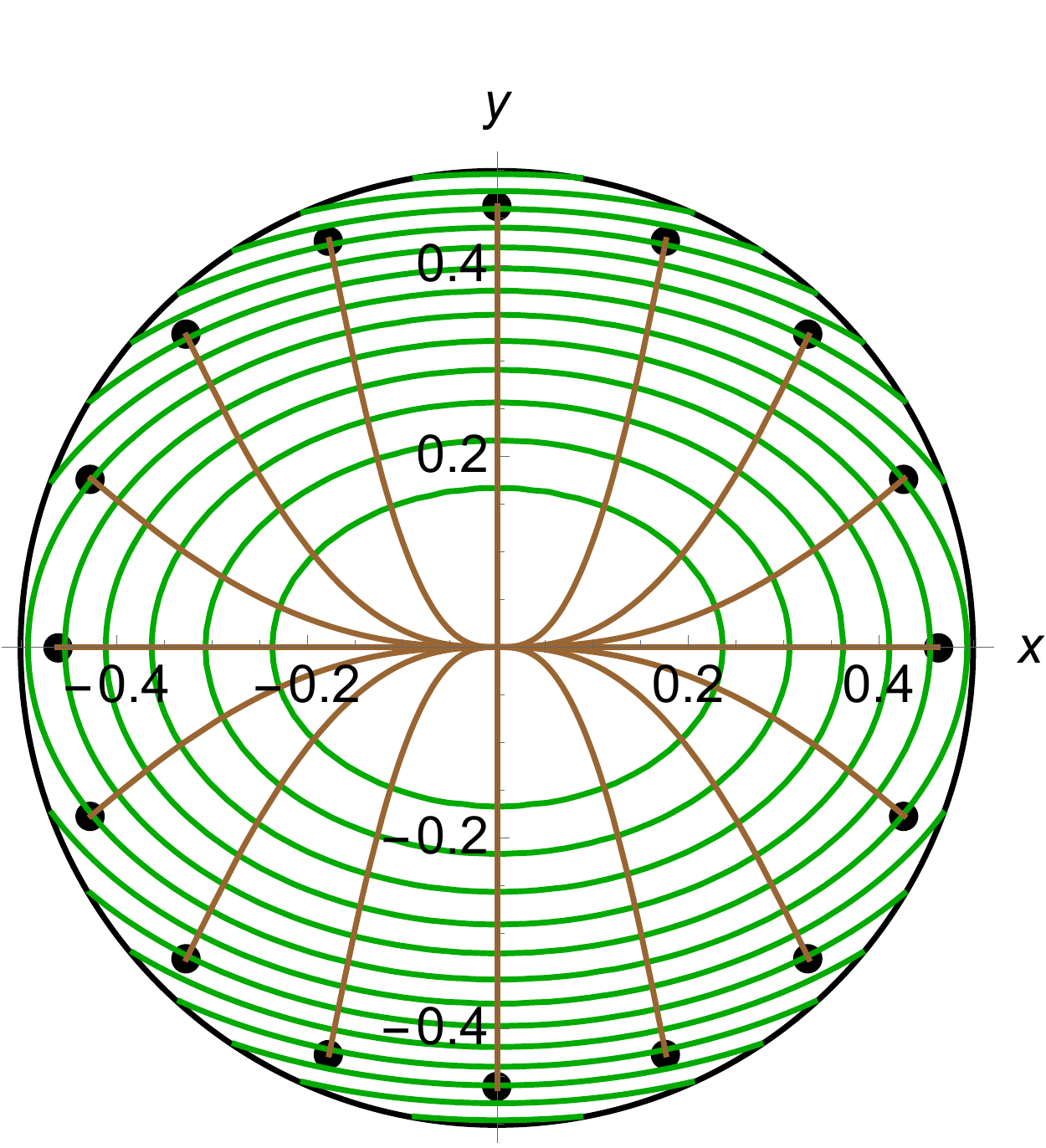} \subcaption{
Local extremum of $V$ for $\beta_\c\!\!=\!0.5$. The dots are initial points $\varphi(0)$.}
\end{minipage}
\caption{Numerically computed orbits of infrared optimal cosmological
curves $\varphi$ of the uniformized model (shown in brown) near an interior
critical point $\c$, superposed over the level lines of $\Phi$ (shown
in green) for two values of $\beta_\c$. Here $x,y$ are principal Cartesian canonical coordinates centered at the critical point. We assume $\omega_\rmax(\c)=0.5$. The initial points
$\varphi(0)$ of these curves are shown as black dots.}
\label{fig:CosmIntCritical}
\end{figure}

Below we compare graphically the effective gradient flow orbits given by solutions of equation \eqref{gradflowc}, depicted in Figure \ref{fig:IntCritical}, to the numerically computed orbits of IR optimal cosmological curves, represented in Figure \ref{fig:CosmIntCritical}.  

\vspace{-2em}
\vspace{-2em}

\section{Brief announcement of further results}

Cosmological curves of two-field models can be approximated by {\em
  mean field curves}, using an approximation technique which is
similar to the mean field approximation of condensed matter
physics. Cosmological mean field approximations admits an elegant
formulation using Ehresmann connections defined on the total space of
the tangent bundle $T\Sigma$, i.e. rank two distributions $\cH\subset
TT\Sigma$ which are complementary to the vertical distribution
$\cV\subset TT\Sigma$ of the fiber bundle $T\Sigma\rightarrow \Sigma$.
The corresponding mean field approximation replaces cosmological {\em
  flow} curves $\gamma=\dot{\varphi}:I\rightarrow T\Sigma$ of the
model with the horizontal lift relative to $\cH$ of curves in $\Sigma$
which satisfy the so-called {\em mean field curve equation} defined by
$\cH$. This amounts to treating as small the components of
$\ddot{\varphi}:I\rightarrow TT\cM$ which are ``orthogonal'' to $\cH$
-- in a sense which can be made precise.

The simplest approximations of this type are induced by the choice of
a {\em special coordinate system} on an open subset of the tangent
bundle $T\Sigma$, i.e. a coordinate system which naturally combines a
coordinate system on the base $\Sigma$ with a coordinate system for
the fibers of $T\Sigma$. In this case, the corresponding Ehresmann
connection is flat and the mean field approximation amounts to
neglecting the first time derivative of the two fiberwise coordinates,
which are thereby being treated as ``slow''. This parallels the logic
of Born-Oppenheimer type approximations in quantum mechanics and
statistical physics, which separate dynamical variables into ``slow''
and ``fast'' and treat the dynamics of slow variables approximately.

In our situation, the first order system of four ODEs which describes
the cosmological equation in a given special coordinate system is
replaced by the algebro-differential system in which the time
derivatives of the fiberwise coordinates are set to zero. This leads
to algebraic consistency conditions for the special coordinates called
{\em mean field equations}, which determine a {\em mean field surface}
inside $T\Sigma$; in many cases, the latter is a semialgebraic
multisection of $T\Sigma$ defined on an open subset of the base
$\Sigma$.

Mean field approximations of this type provide a very general
procedure for extracting approximants of cosmological curves in
various regimes, where the regime of interest is defined by the choice
of fiberwise coordinates that one wishes to treat as ``slow''. Since
fiberwise coordinates are pairs of basic observables of the
cosmological system which are functionally independent on an open
subset of $T\Sigma$, each such regime is determined by the choice of a
pair of locally independent on-shell cosmological observables.  As we
show in forthcoming work, a careful study of natural on-shell
observables of two-field cosmological models provides interesting
candidates for such fiberwise coordinates on $T\Sigma$, thus leading
to natural mean field approximation schemes which have direct physics
significance.  The latter can be applied to any two-field model and in
particular to tame hyperbolizable models.

One such mean field approximation is the so-called {\em adapted
  approximation}. This uses the fiberwise coordinates on $T\Sigma$
which are given by the projections of $\dot{\varphi}$ on the direction
of the vector $\grad_G \Phi$ and on its positive normal direction and
provides a mathematical refinement of the proposal of
\cite{Bjorkmo}. One can also consider the {\em roll-turn
  approximation}, which takes as fiberwise coordinates the {\em
  on-shell} second slow roll parameter and turning rate. Finally, one
can consider the {\em slow roll rate approximation}, which uses the
first and second {\em on shell} slow roll parameters as fiberwise
coordinates.

Another kind of approximation which can be considered for two-field
cosmological models is the so-called {\em angular approximation},
which arises by neglecting the first two time derivatives of the
radial variable in a given semigeodesic coordinate system
$(r,\theta)$. This provides approximants on each semigeodesic
coordinate patch, whose accuracy can be characterized theoretically
and computed numerically.

In forthcoming work, we study the approximations mentioned above for
tame two-field cosmological models and the corresponding error terms,
which play an important role when ascertaining their accuracy (see
\cite{AL}).  In particular, this serves as a test of various proposals
made previously in the literature. Moreover, we compare these
approximations with the IR approximation studied in \cite{grad} (and
summarized above) near interior critical points and near ends of
$\Sigma$, determining the regimes within which the various
approximations are accurate.

\section{Conclusions}

We studied the first order IR behavior of tame hyperbolizable
two-field cosmological models by analyzing the asymptotic form of the
gradient flow orbits of the classical effective scalar potential $V$
with respect to the uniformizing metric $G$ near all interior critical
points and ends of $\Sigma$. We showed that the IR behavior of tame
hyperbolizable two-field cosmological models is characterized by a
finite set of parameters associated to their ends and interior
critical points. Comparing with numerical computations, we found that
the first order IR approximation is already quite good for all
interior critical points and all ends except for cusps, for which one
must consider higher order corrections in the IR expansion in order to
obtain a good approximation. Our results characterize the IR
universality classes of all tame hyperbolizable two-field models in
terms of geometric data extracted from the asymptotic behavior of the
effective scalar potential and uniformizing metric.

Since the Morse assumption on the extended potential determines its
asymptotic form near all points of interest on $\hSigma$, we could
derive closed form expressions for the asymptotic gradient flow which
describes the corresponding infrared phases of such models in the
sense of \cite{ren}. In particular, we found that the asymptotic
gradient flow of $(\Sigma,G,V)$ near each end which is a critical
point of the extended potential can be expressed using the incomplete
gamma function of order two and certain constants which depend on the
type of end under consideration and on the principal values of the
extended effective potential $\hV$ at that end. We also found that
flaring ends which are not critical points of $\hV$ act like
fictitious but non-standard stationary points of the effective
gradient flow. While the local form near the critical points of $V$ is
standard (since they are hyperbolic stationary points
\cite{Palis,Katok} of the cosmological and gradient flow), the
asymptotic behavior near Freudenthal ends is exotic in that some of
the ends act like fictitious stationary points with unusual
characteristics.

We compared these results with numerical computations of cosmological
curves near the points of interest. We found particularly interesting
behavior near cusp ends, around which generic cosmological
trajectories tend to spiral a large number of times before either
``falling into the cusp'' or being ``repelled'' back toward the
compact core of $\Sigma$ along principal geodesic orbits determined by
the classical effective potential $V$.  In particular, cusp ends lead
naturally to ``fast turn'' behavior of cosmological curves.

\acknowledgments{\noindent This article was supported by grant
  PN 19060101/2019-2022. The authors also acknowledge this paper 
as part of their contribution within the COST Action COSMIC WISPers CA21106, 
supported by COST (European Cooperation in Science and Technology).}

\end{document}